\documentclass[preprint]{elsarticle}

\setcitestyle{super}
\newcommand{\onlinecite}[1]{\hspace{-1 ex} \nocite{#1}\citenum{#1}} 

\biboptions{sort&compress}

\usepackage{amsmath}
\usepackage{amssymb}
\usepackage{amstext}
\usepackage{amsbsy}
\usepackage{amsthm}
\usepackage{graphicx}
\usepackage{epstopdf}
\usepackage{epsfig}
\usepackage{hyperref}
\usepackage{color}
\usepackage{multirow}
\usepackage{natbib}

\newcommand {\B}[1]{\textcolor{blue}{#1}}
\newcommand{\Fig}[1]{Fig.~\ref{#1}}
\newcommand{\Ref}[1]{Ref.~\onlinecite{#1}}

\newcommand{\Eq}[1]{Equation~(\ref{#1})}
\newcommand{\nn}{\nonumber\\}

\newcommand{\Rmnum}[1]{\expandafter\@slowromancap\romannumeral #1@}
\newcommand{\ie}{{\emph{i.e.~}}}

\newcommand{\be}{\begin{eqnarray}}
\newcommand{\ee}{\end{eqnarray}}
\newcommand{\beq}{\begin{equation}}
\newcommand{\eeq}{\end{equation}}
\newcommand{\bpm}{\begin{pmatrix}}
\newcommand{\epm}{\end{pmatrix}}
\newcommand{\bal}{\begin{aligned}}
\newcommand{\eal}{\end{aligned}}
\newcommand{\<}{\langle}
\renewcommand{\>}{\rangle}
\newcommand{\Tr}{{\rm Tr}}

\newcommand{\p}{\partial}
\renewcommand{\t}[1]{{\tilde #1}}

\newcommand{\ra}{\rightarrow}

\newcommand{\e}{\epsilon}

\newcommand{\w}{{\omega}}
\newcommand{\s}{{\sigma}}

\newcommand \ti[1]{}

\newcommand{\dg}{\dagger}
\newcommand{\ket}[1]{|{#1}\rangle}
\newcommand{\ud}{\mathrm{d}}
\newcommand{\dd}[3][]{\ud^{#1} #2 / \ud #3^{#1}}
\newcommand{\ch}{\chi}

\renewcommand{\t}{\tau}

\begin{document}

\author[ucb]{Lokman Tsui}
\ead{lokman@berkeley.edu}

\author[ucb]{Yen-Ta Huang}
\ead{yenta.huang@berkeley.edu}

\author[sf]{Hong-Chen Jiang}
\ead{hcjiang@stanford.edu}

\author[ucb,lbl]{Dung-Hai Lee\corref{cor1}}
\ead{dunghai@berkeley.edu}
\cortext[cor1]{Corresponding author}

\address[ucb]{Department of Physics, University of California, Berkeley, California 94720, USA}
\address[lbl]{Materials Sciences Division, Lawrence Berkeley National Laboratories, Berkeley, California 94720, USA}
\address[sf]{Stanford Institute for Materials and Energy Sciences, SLAC National Accelerator Laboratory and Stanford University, 2575 Sand Hill Road, Menlo Park, CA 94025, USA}

\date{\today}

\title{The phase transitions between $Z_n\times Z_n$ bosonic topological phases in 1+1 D, and a constraint on the central charge for the critical points between bosonic symmetry protected topological phases
}


\begin{abstract}
The study of continuous phase transitions triggered by spontaneous symmetry breaking has brought 
revolutionary ideas to physics.  Recently, through the discovery of symmetry protected topological phases, it is realized that continuous quantum phase transition can also occur between states with the same symmetry but different topology. Here we study a specific class of such phase transitions in 1+1 dimensions -- the phase transition between bosonic topological phases protected by $Z_n\times Z_n$. We find in all cases the critical point possesses two gap opening relevant operators: one leads to a Landau-forbidden symmetry breaking phase transition and the other to the topological phase transition. We also obtained a constraint on the central charge for general phase transitions between symmetry protected bosonic topological phases in 1+1D.   
\end{abstract}

\maketitle

\section{Introduction and the outline}

Last five years witnessed a fast progress in the understanding of a new type of quantum disordered states -- symmetry protected topological states (SPTs)\cite{Schnyder2008,Kitaev2009,Chen2011}. 
These states exhibit a full energy gap in closed (i.e., boundary-free) geometry and exhibit the full symmetry of the Hamiltonian. 
However, these states are grouped into different ``topological classes'' such that it is not possible to cross from one topological class to another without closing the energy gap while preserving the symmetry.   Our goal is to understand the difference (if any) between the traditional Landau type and this new kind of ``topological'' phase transitions.\\ 

Because the Landau-type phase transitions are triggered by the fluctuations of bosonic order parameters over space-time, to minimize the obvious difference we focus on the phase transitions between bosonic SPT phases\cite{Chen2011}. Hence we do not address the phase transition between fermionic topological insulators or superconductors\cite{Schnyder2008,Kitaev2009}. Moreover, to make everything as concrete as possible we shall focus on one space dimension and to topological phase transitions which have dynamical exponent equal to one (hence can be described by conformal field theories (CFTs)). 

We spend most of the space describing the study of a specific class of such phase transitions -- the phase transition between bosonic SPTs protected by $Z_n\times Z_n$.
Here we combine a blend of analytic and numerical methods to arrive at a rather complete picture for such critical points. From studying these 
phase transitions we observe an interesting fact, namely whenever the transition is {\it direct} (i.e., when there are no intervening phases) and {\it continuous} the central charge ($c$) of the CFT is always greater or equal to one. Near the end of the paper, we obtain a constraint on the central charge for 
CFTs describing bosonic SPT phase transitions: namely, $c\ge 1$. Therefore, none of the best known ``minimal models\cite{Belavin1984}'' can be the CFT for bosonic SPT phase transitions!\\

According to the group cohomology classification\cite{Chen2011}, in one space dimension, the group $Z_n\times Z_n$ protects $n$ different topological classes of SPTs. If we ``stack'' a pair of SPTs (which can belong to either the same or different topological class) on top of each other and turn on all symmetry allowed interactions, a new SPT will emerge to describe the combined system. An abelian (cohomology) group $H^2(Z_n\times Z_n,U(1))=Z_n$ (here the superscript ``2'' refers to the space-time dimension) classifies the SPT phases and describes the stacking operation. Here each topological class is represented by an element (i.e., $0,...,n-1$) of $H^2(Z_n\times Z_n,U(1))=Z_n$ and the ``stacking'' operation is isomorphic to the $mod(n)$ addition of these elements. \\ 

To understand the phase transitions between different classes of SPTs it is sufficient to focus on the transition between the trivial state (which corresponds to the ``0'' of  $Z_n$) and the non-trivial SPT corresponding to the ``1'' of $Z_n$. The transition between phases correspond to other adjacent elements of $Z_n$, e.g., $(m,m+1)$, will be in the same universality class as that between $(0,1)$. Transitions between ``non-adjacent'' topological classes will generically spit into successive  transitions between adjacent classes. \\

There are 11 sections in the main text. In these sections we restrain from heavy mathematics, i.e., we simply state the main results and provide simple arguments. There are 6 appendices where mathematical details can be found. The outline of this paper is the follows. In section 2 we present the exactly solvable fixed point hamiltonians for the trivial and non-trivial $Z_n\times Z_n$ protected SPT phases. In section 3 we present a hamiltonian that interpolates between the fixed point hamiltonians in section 2. A single parameter tunes this hamiltonian through the SPT phase transition. Section 4 introduces a non-local transformation that maps the hamiltonian in section 3 to that of two $n$-state clock models with spatially twisted boundary condition and Hilbert space constraint. In particular, at criticality, we show that the partition function of the transformed hamiltonian corresponds to an ``orbifolded'' $Z_n\times Z_n$ clock model. In section 5 we discuss the effects of orbifolding on the phases of the clock model and show the results are consistent with what one expect for the SPT phases. Section 6 gives the phase diagram of the hamiltonian given in section 3. In section 7 we show that from the point of view of the orbifolded clock model the SPT transition corresponds to a Landau forbidden transition. In section 8 we present the conformal field theories for the SPT phase transitions discussed up to that point. Section 9 presents our numerical density matrix renormalization group results. We compare these results with the prediction of section 8. Section 10 presents the argument that the central charge of the CFTs that describe SPT phase transitions must be greater or equal to one. Finally, section 11 is the conclusion. \\

In appendix A, we provide a brief review of the key ingredients of the $1+1$D group cohomology, namely, the notions of cocycles and projective representations. After that, we show how to use cocycles to construct solvable fixed point SPT hamiltonians. Appendix B summarizes the non-local transformation that maps the hamiltonian in section 3 of the main text to that of two $n$-state clock models with spatially twisted boundary condition and Hilbert space constraint. In appendix C we show that the partition function associated with the hamiltonian in appendix B (and section 3 of the main text) corresponds to that of ``orbifolded'' $Z_n\times Z_n$ clock model. Appendices D,E,F present the modular invariant partition functions of the orbifold $Z_2\times Z_2$, $Z_3\times Z_3$ and $Z_4\times Z_4$ clock models, respectively. In these appendices, we examine the primary scaling operator content of the modular invariant conformal field theory. In addition, we study the symmetry transformation properties of various Verma modules and the scaling dimension of primary scaling operators, particularly that of the gap opening operator. Appendix G summarizes the details of the density matrix renormalization group calculation. Finally, in appendix H we briefly review the symmetry of the minimal model conformal field theories.

\section{Exactly solvable ``fixed point'' Hamiltonians for the SPTs\\}

Each SPT phase is characterized by an exactly solvable ``fixed point'' Hamiltonian. 
In appendix \ref{appdx:znznh} we briefly review the construction of these  Hamiltonians using the ``cocycles'' 
associated with the cohomology group\cite{Ran2012,Lokman2015}. For the case relevant to our discussion the following lattice Hamiltonians can be derived\cite{Santos2015} so that its ground state belong to the ``0'' and ``1'' topological classes of $H^2(Z_n\times Z_n,U(1))=Z_n$ 
\begin{align}
H_0&=-\sum\limits_{i=1}^{N} (M_{2i-1}+M_{2i} + h.c.)\nn
H_1&=-\sum\limits_{i=1}^{N} (R^{\dagger}_{2i-2} M_{2i-1} R_{2i}+R_{2i-1} M_{2i} R^{\dagger}_{2i+1}+h.c.) \label{eq:znznh}
\end{align}
These Hamiltonians are defined on 1D rings consisting of $N$ sites.
For each site labeled by $i$ the local Hilbert space is spanned by 
 $|g_{2i-1},g_{2i}\>:=|g_{2i-1}\>\otimes|g_{2i}\>$ where $(g_{2i-1},g_{2i})\in Z_n\times Z_n$ with $g_{2i-1},g_{2i}=0,1,...,n-1$. The total Hilbert space is the tensor product of the local Hilbert space for each site. For the convenience of future discussions from now on we shall refer to $(2i-1,2i)$ as defining  a ``cell'', and call  $|g_{2i-1}\>$ and $|g_{2i}\>$ as basis states defined for ``site'' $2i-1$ and  $2i$.
The operators $M_j$ and $R_j$ in \Eq{eq:znznh} are defined by
\be
&&M_j \ket{g_j}:=\ket{g_j+1}~~ {\rm mod~}n, ~~{\rm and}\nn
&&R_j \ket{g_j}:=\eta_n^{g_j} \ket{g_j}~~{\rm where~}\eta_n=e^{i2\pi/n}.
\label{mr}
\ee 
From \Eq{mr} we deduce the following commutation relation between $M$ and $R$:
\be
 &&R_j R_k=R_k R_j\nn
 &&M_j M_k=M_k M_j\nn
 &&R_j M_k = \eta_n^{\delta_{jk}} M_k R_j.
 \label{com}
\ee
Due to this commutation relation, it can be checked that the $n\times n$ matrices associated with $M_j$ and $R_j$ form a {\it projective} representation of the $Z_n\times Z_n$ group multiplication law (see appendix \ref{appdx:projrep} for the definition of projective representations).
Finally periodic boundary condition is imposed on \Eq{eq:znznh} which requires \be g_{2N+1}=g_1, ~~{\rm and}~~g_{2N+2}=g_2.\label{pbc}\ee Under these definitions \Eq{eq:znznh} is invariant under the global $Z_n\times Z_n$ group generated by \be\prod_{i=1}^N M_{2i-1}~~{\rm and}~~\prod_{i=1}^N M_{2i}.\label{sym1}\ee \\

The form of Hamiltonians given in \Eq{eq:znznh} is quite asymmetric between $M$ and $R$. We can make it more symmetric by performing the following unitary transformation on the local cell basis as follow 
\begin{align*}
\ket{g_{2i-1},g_{2i}}\ra U\ket{g_{2i-1},g_{2i}}=\frac{1}{\sqrt{n}}\sum_{g_{2i}^\prime=0}^{n-1}\eta_n^{(g_{2i-1}-g_{2i})g'_{2i}}\ket{g_{2i-1},g_{2i}^\prime}.
\end{align*}
This results in the following transformations of the operators in \Eq{eq:znznh}
\be
 &&U^\dagger M_{2i-1} U = M_{2i-1} M_{2i}\nn
 &&U^\dagger M_{2i} U = R^{\dagger}_{2i-1} R_{2i} \nn
 &&U^\dagger R_{2i-1} U = R_{2i-1} \nn
 &&U^\dagger R_{2i} U = M^{\dagger}_{2i}. 
\label{transf}
\ee
It is straightforward to show that after these transformations  the new operators obey the same commutation relation as \Eq{com}. Moreover, it can also be shown that $R$ obeys the same boundary condition, namely,
$R_{2N+1}=R_1$ and $R_{2N+2}=R_2$. In addition, it is also straightforward to show that under \Eq{transf} the generators of the $Z_n\times Z_n$ group become 
\be
&&U^\dagger \left(\prod_{i=1}^N M_{2i-1}\right) U=\prod_{j=1}^{2N} M_j~~~{\rm and}~~~U^\dagger \left(\prod_{i=1}^N M_{2i}\right) U=\prod_{j=1}^{2N} R_j^{(-1)^j}.
\label{gen}
\ee
Thus alternating ``site'' carries the projective and anti-projective representation of
$Z_n\times Z_n$.\\

\begin{figure}[h!]
\centering
\includegraphics[width=8cm]{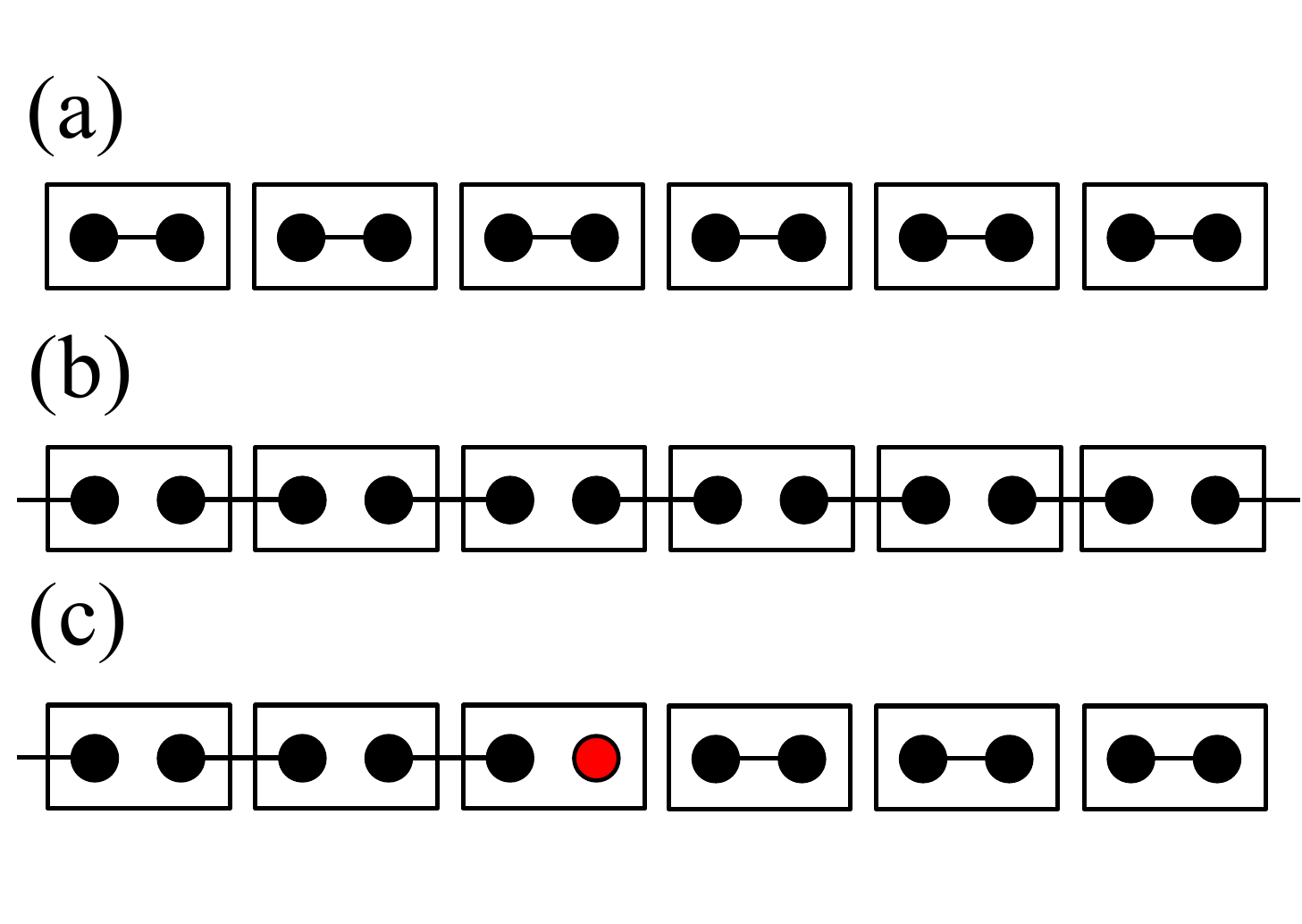}
\caption{(Color online) (a) $H_0$ couples states associated with the same cell (each cell is represented by the rectangular box). (b) $H_1$ couples states associated with adjacent cells. Each pair of black dots in a rectangle represents the sites in each cell. They carry states $|g_{2i-1}\>$ and $|g_{2i}\>$ which form a projective representation of $Z_n\times Z_n$.  Each link represents a coupling term in the Hamiltonian \eqref{eq:hdimer}. (c) Hamiltonian describing the interface between the two SPTs each being the ground state of $H_0$ and $H_1$. It is seen that there is a leftover site (highlighted in red) transforming projectively at the interface.}
\label{fig:chain}
\end{figure}

Under \Eq{transf} the Hamiltonian $H_0$ and $H_1$  become
\begin{align} 
H_0&=-\sum\limits_{i=1}^{N} (M_{2i-1}M_{2i}+R_{2i-1}R_{2i}^{\dagger} + h.c.) \nn
H_1&=-\sum\limits_{i=1}^{N} (M_{2i} M_{2i+1} + R_{2i} R^{\dagger}_{2i+1}+h.c.) \label{eq:hdimer}
\end{align}
These Hamiltonians are pictorially depicted in \Fig{fig:chain}(a,b). Note that while $H_0$ (\Fig{fig:chain}(a)) couples sites within the same cell, $H_1$ couples sites belong to adjacent cells (\Fig{fig:chain}(b)).
Because both $H_0$ and $H_1$ consist of decoupled pairs of sites (the coupling terms associated with different pairs commute with one another)  they can be exactly diagonalized. The result shows a unique ground state with a fully gapped spectrum for both $H_0$ and $H_1$. 
Using \Eq{gen} it is simple to show that the ground states are invariant under $Z_n\times Z_n$.\\ 

The fact that $H_0$ and $H_1$ describe inequivalent SPTs can be inferred by forming an interface of $H_0$ and $H_1$ as shown in \Fig{fig:chain}(c). A decoupled site (red) emerges. Localizing on this site there are degenerate gapless excitations carrying a projective representation of the $Z_n\times Z_n$\cite{Cho2016}. The fact that gapless excitations must exist at the interface between the ground states of $H_0$ and $H_1$ attests to that fact that these states belong to inequivalent topological classes of  $H^2(Z_n\times Z_n,U(1))=Z_n$.  

\section{An interpolating Hamiltonian describing the phase transition between $Z_n\times Z_n$ SPTs}

To study the phase transition between the ground state of $H_0$ and the ground state of $H_1$
we construct the following Hamiltonian which interpolates between $H_0$ and $H_1$ as follows
\be
&&H_{01}(\lambda)=(1-\lambda) H_0 + \lambda H_1\nn
&&=-(1-\lambda)\sum\limits_{i=1}^{N} (M_{2i-1}M_{2i}+R_{2i-1}R_{2i}^{\dagger})
-\lambda\sum\limits_{i=1}^{N} (M_{2i} M_{2i+1} + R_{2i} R^{\dagger}_{2i+1})\nn&&+h.c. \label{eq:hlambda}
\ee
With both $H_0$ and $H_1$ present the Hamiltonian given in \Eq{eq:hlambda} is no longer easily solvable. However, in the following, we present analytic results showing (1) for $2\le n\le 4$ the phase transition occurs at $\lambda=1/2$, (2) the central charge, the conformal field theory and its associated primary scaling operators at the phase transitions. 
For $n\ge 5$ there is a gapless phase centered around $\lambda=1/2$ hence the phase transition is not direct. 
Moreover for the interesting case of $n=3$ we will present the numerical density matrix renormalization group results which confirm our analytic solution.

\section{Mapping to ``orbifold'' $Z_n\times Z_n$ clock chains\\}\label{dual main}

In appendix \ref{appdx:of}
 we show that  \Eq{eq:hlambda} can be mapped onto a  $Z_n\times Z_n$ clock model with spatially twisted boundary condition and a Hilbert space constraint. In appendix \ref{appdx:orbifold} we further show that these amount to ``orbifolding'',

 The mapping is reminiscent of the duality transformation in a single $Z_n$ clock model.  The mapping is achieved via the following transformations:
\be
&&R^{\dagger}_{j-1}R_{j}=\widetilde{M}_j~~{\rm for}~~j=2 ...2N,~~ R^{\dagger}_{2N}R_{1}= \widetilde{M}_{1}
\nn&&{\rm and}~~M_{j}= \widetilde{R}^{\dagger}_{j} \widetilde{R}_{j+1}~{\text{for all}}~j.
\label{dualtrf}
\ee
After the mapping, the Hamiltonian in \Eq{eq:hlambda} is transformed to
\be
&&H_{01}(\lambda)=H_{\rm even}(\lambda)+H_{\rm odd}(\lambda)\nn
&&H_{\rm even}(\lambda)=-\sum\limits_{i=1}^{N}\left[(1-\lambda)\widetilde{M}_{2i}+\lambda \widetilde{R}^{\dagger}_{2i}\widetilde{R}_{2i+2} \right]+ h.c.\nn
&&H_{\rm odd}(\lambda)=-\sum\limits_{i=1}^{N}\left[\lambda\widetilde{M}_{2i-1}+(1-\lambda) \widetilde{R}^{\dagger}_{2i-1}\widetilde{R}_{2i+1} \right]+ h.c.
\label{dualH}
\ee 
Here $\widetilde{M}$ and $\widetilde{R}$ obey the same commutation relations as $M$ and $R$ in \Eq{com}. 
\\

\Eq{dualH} is the quantum Hamiltonian for two $Z_n$ clock models\cite{Jose1977}, one defined on the even and one on the odd sites, respectively.
However, generated by the mapping, \Eq{dualH} is 
supplemented with a twisted spatial boundary condition and a constraint: 
\be
&&{\rm Boundary~condition:}~~\widetilde{R}_{2N+1}:= \widetilde{B}\widetilde{R}_{1}\text{~~and~~} \widetilde{R}_{2N+2}:= \widetilde{B}\widetilde{R}_{2},\label{bct}\ee 
\be 
{\rm Constraint:}~~\prod _{i=1}^{2N} \widetilde{M}_i=1.\label{tcon}
\ee Here $\widetilde{B}$ is an operator that commutes with all the $\widetilde{R}$s and $\widetilde{M}$s. The eigenvalues of $\widetilde{B}$ are $\tilde{b}=1,\eta_n,...,\eta_n^{n-1}$ (recall that $\eta_n=e^{i 2\pi/n}$).
In terms of the transformed variables, the generators of the original $Z_n\times Z_n$ group are given by
\be
\widetilde{B}\text{~~and~~}\prod_{j=1}^N \widetilde{M}_{2j}. \label{dualsym}
\ee
The spatially twisted boundary condition \Eq{bct} and the constraint \Eq{tcon} (which turns into a time direction boundary condition twist in the path integral representation of the partition function) execute the ``orbifolding'' 
(see later).
\\

By swapping the even and odd chains \Eq{dualH} 
exhibit the $$\lambda\leftrightarrow (1-\lambda)$$ duality.
This implies the self-dual point at $\lambda=1/2$ is special. In particular, {\it if}  there is a single critical point as a function of $\lambda$, it must occur at $\lambda=1/2$. Incidentally, if we put aside \Eq{bct} and \Eq{tcon}, $\lambda=1/2$ is where each of the clock chains in \Eq{dualH} becomes critical.\\ 

As we will show later the effects of \Eq{bct} and \Eq{tcon} (i.e., orbifold) is to change the primary scaling operator content of the critical CFT from that of the direct product of two $Z_n$ clock models. However they do not jeopardize the criticality, nor do they change the central charge. We shall return to these more technical points later.  At the meantime let's first study the effects of \Eq{bct} and \Eq{tcon} on the phases.
\\

\section{The effect of orbifold on the phases}\label{phases}

Knowing the behavior of the single $Z_n$ clock chain, \Eq{dualH} suggests for $\lambda<1/2$  the odd-site chain will spontaneously break the $Z_n$ symmetry while the even chain remains disordered. The ground state will lie in the $\tilde{b}$ (the eigenvalue of $\tilde{B}$) $=1$ sector on account of the twisted boundary condition. For $\lambda>1/2$  the behaviors of the even and odd chains exchange, and the ground state remains in the $\tilde{b}=1$ sector. On the surface, such symmetry breaking should lead to ground state degeneracy which  is inconsistent with the fact that both SPTs (for $\lambda<1/2$ and $\lambda>1/2$) should have unique groundstate.\\

This paradox is resolved if we take into account of the constraint in \Eq{tcon}. For simplicity let's look at the limiting cases. For $\lambda=0$ the ground state of \Eq{dualH} is
\be
|g,g,...,g\rangle_{\rm odd}\otimes |p,p,...,p\rangle_{\rm even}\otimes |\tilde{b}=1\rangle\ee 
where 
$g=0,...,n-1$. Here the ``paramagnet state'' $|p\rangle$ for each site is defined as
\be
|p\rangle:={1\over\sqrt{n}}\left(|0\rangle+|1\rangle+...+|n-1\rangle\right).\ee As expected, such ground state is $n$-fold degenerate and it does not satisfy the constraint of \Eq{tcon}. However, if we form the symmetric superposition of the odd-site symmetry breaking states
\be 
\left(\frac{1}{\sqrt{n}}\sum_{g=0}^{n-1}|g,g,...,g\rangle_{\rm odd}\right)\otimes |p,p,...,p\rangle_{\rm even}\otimes |\tilde{b}=1\rangle
\label{gnd}\ee 
the constraint is satisfied and the state is
non-degenerate. Obviously, \Eq{gnd} is invariant under the $Z_n\times Z_n$ generated by \Eq{dualsym}. Although \Eq{gnd} is non-degenerate, the two-point correlation function $\langle \widetilde{R}_{2j+1} \widetilde{R}^\dagger_{2k+1}\rangle$ still
shows long-range order.
Almost exactly the same arguments, with odd and even switched, apply to the $\lambda=1$ limit. The only difference is instead of observing $|p,p,...,p\rangle_{\rm even}$ being invariant under the action of  $\prod_{j=1}^N \widetilde{M}_{2j}$ we need to observe 
that $\left(\frac{1}{\sqrt{n}}\sum_{g=0}^{n-1}|g,g,...,g\rangle_{\rm even}\right)$ is invariant. As $\lambda$ deviates from the limiting values, so long as it does not cross any phase transition the above argument should remain qualitatively unchanged. In this way we understand the effects of \Eq{bct} and \Eq{tcon} on the phases.  \\

\section{The phase diagram}\label{phase diag}

\begin{figure}[h!]
\centering
\includegraphics[scale=0.45]{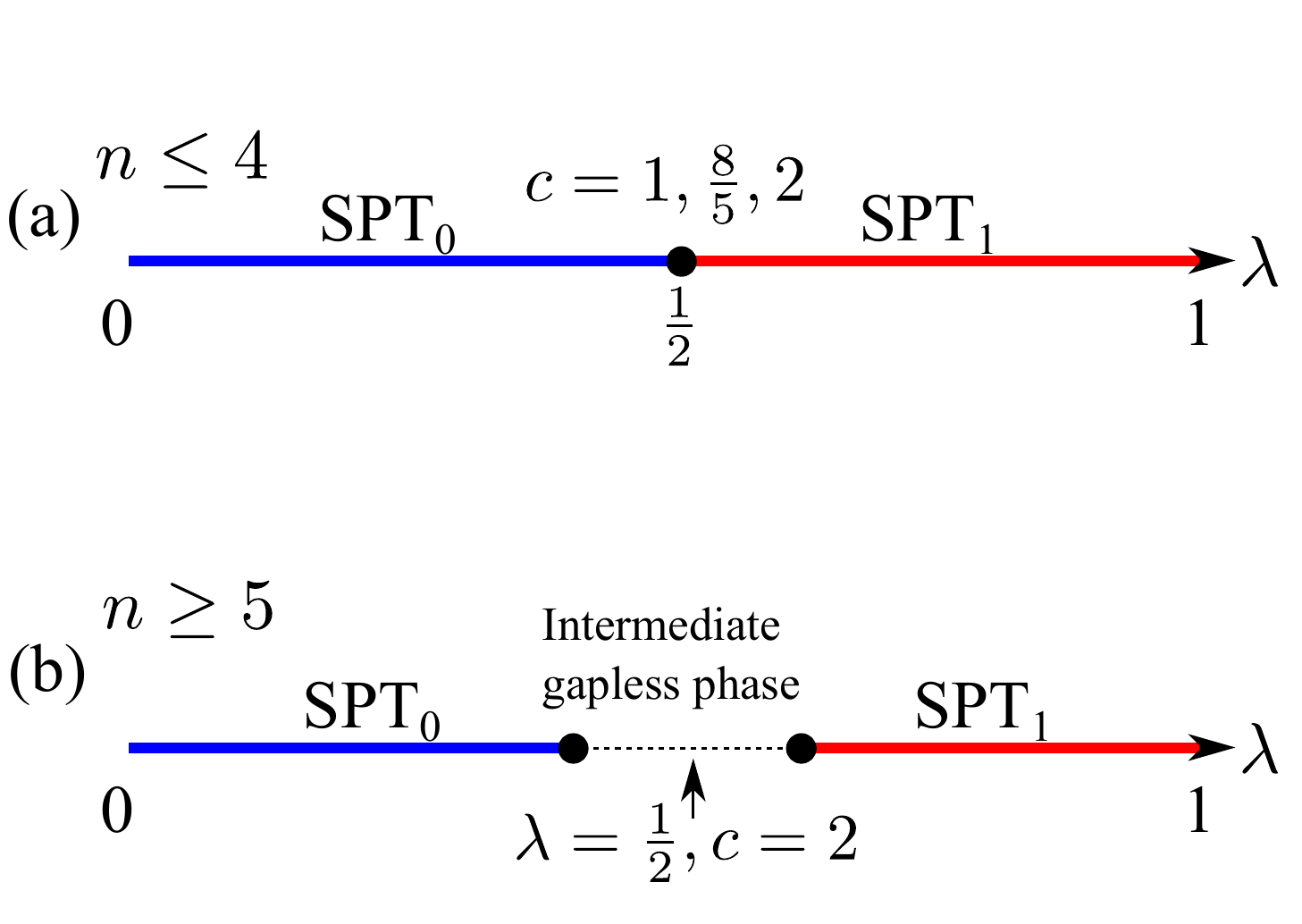}
\caption{(Color online) Phase diagram for \eqref{eq:hlambda}, which linearly interpolates between the  fixed point hamiltonians of $Z_n\times Z_n$ SPT phases. Red and blue mark the non-trivial and trivial SPTs respectively. (a) For $n\leq4$, a second-order transition occurs between the two SPT phases, and the central charge takes values of 1, $\frac{8}{5}$ and 2  for $n=2,3,4$, respectively. (b) For $n\geq5$, a gapless phase intervenes between the two SPT phases. The entire gapless phase has  central charge $c=2$.}
\label{fig:znphase}
\end{figure}

Since upon orbifolding the phases of the decoupled $Z_n\times Z_n$ clock models seamlessly evolve into the SPT phases we shall construct that phase diagram using what's know about the phase structure of the clock model. It is known that a single $Z_n$ clock chain shows an order-disorder phase transition at a single critical point for $n\le 4$, while there is an intermediate gapless phase for $n\ge 5$  we conclude the phase diagram is shown in \Fig{fig:znphase}(a,b). 
Since our goal is to study the continuous phase transition between SPTs we focus on $n\le 4$.

\section{SPT transitions as ``Landau-forbidden'' phase transitions}

According to Landau's rule, transitions between phases whose symmetry groups do not have subgroup relationship should generically be first order. {\it Continuous} phase transitions between such phases are regarded as ``Landau forbidden'' in the literature.\\
 
As discussed earlier, in terms of the orbifolded $Z_n\times Z_n$ clock chains, the two phases on either side of the SPT phase transition correspond to the breaking of the $Z_n$ symmetry in one of the clock chain but not the other. In the following, we elaborate on this statement.\\

For $\lambda<1/2$ although the ground state in \Eq{gnd} is non-degenerate, the two-point correlation function $\langle \widetilde{R}_{2j+1} \widetilde{R}^\dagger_{2k+1}\rangle$
shows long-range order. When the odd and even chains are switched the same argument applies to the $\lambda>1/2$ limit. 
If we define 
\be
Q_{\rm even}=\prod_{j=1}^N \widetilde{M}_{2j} \text{~and~} Q_{\rm odd}=\prod_{j=1}^N \widetilde{M}_{2j-1}\ee
it is easy to show that equations \eqref{dualH}, \eqref{bct} and \eqref{tcon}  commute with them, hence the $Z^\prime_n\times Z^\prime_n$ group
they generate are also the symmetry of the problem. However it is important not to confuse $Z^\prime_n\times Z^\prime_n$ with the original $Z_n\times Z_n$ group (which is generated by \Eq{dualsym}). \\

With respect to the $Z^\prime_n\times Z^\prime_n$ symmetry the two phases (realized for $\lambda<1/2$ and $\lambda>1/2$) breaks two different $Z^\prime_n$ factors, hence the symmetry groups of the two phases have no subgroup relationship, thus if a continuous phase transition between them exists it is a Landau forbidden transition. In fact, it is the original $Z_n\times Z_n$ symmetry that ``fine tunes'' the system to realize such non-generic continuous phase transition.

\section{The CFT at the SPT phase transition for $n=2,3,4$}\label{cft}

It is known that the central charge of the CFT describing the criticality of a {\it single} $Z_n$ clock chain is  $c=1/2,4/5,1$ for $n=2,3,4$. Thus the central charge of the CFT describing the simultaneous criticality of two decoupled $Z_n$ clock chains should be $c=1,8/5,2$ for $Z_2\times Z_2$, $Z_3\times Z_3$ and $Z_4\times Z_4$. This is summarized in 
Table I.
\begin{table}
\begin{center}
\begin{tabular}{|c|c|}
\hline
Symmetry group	&Central charge\\
\hline
$Z_2\times Z_2$				&1\\
\hline
$Z_3\times Z_3$			&8/5\\
\hline
$Z_4\times Z_4$    &2\\
\hline
\end{tabular}
\caption{The central charges associated with the critical point of the $Z_n\times Z_n$ SPT phase transitions for $n=2,3,4$.}
\label{c}
\end{center}
\end{table} 
\\

Of course, we do not have two decoupled clock chains. The spatial boundary condition twist (\Eq{bct}) and the constraint (\Eq{tcon}), namely the orbifolding, couples the two chains together. The purpose of this section is to address the effects of  orbifolding on the criticality of the two decoupled chains.\\

Let's start with the conformal field theory of a single $Z_n$ clock chain. The partition function of such CFT on a torus is given by 
\be
Z(q)=\sum_{a,b}\chi_a(q) M_{ab}\bar{\chi}_b(\bar{q}).\label{mab}\ee Here the indices $a,b$ labels the Verma modules. Each Verma module is spanned by states associated with a primary scaling operator and its descendants through the operator-state correspondence. Each Verma module carries an irreducible representation of the conformal group. The parameter $q$ in \Eq{mab} is equal to $e^{2\pi i \tau}$, where $\tau$ is the modular parameter of the spacetime torus (see \Fig{fig:tau}). $\chi_a(q)$ and $\bar{\chi}_b(\bar{q})$ are, respectively, the partition function associated with the holomorphic Verma module $a$ and antiholomorphic Verma module $b$. The matrix $M_{ab}$ has non-negative integer entries.  
\\

The partition function of the two decoupled $Z_n$ clock chains that are simultaneously critical is given by \be
Z(q)\times Z(q)=\sum_{a,b,c,d}\chi_a(q)\chi_c(q)(M_{ab}M_{cd})\bar{\chi}_b(\bar{q})\bar{\chi}_d(\bar{q}).\ee  It turns out that the effect of orbifold is to change \be M_{ab}M_{cd}\ra N_{(a,c);(b,d)}\ee ($N_{(a,c);(b,d)}$ is a different non-negative integer matrix) so that
\be Z_{\rm orbifold}(q)=\sum_{a,b,c,d}\chi_a(q)\chi_c(q)N_{(a,c);(b,d)}\bar{\chi}_b(\bar{q})\bar{\chi}_d(\bar{q}).\ee In particular, $N_{(1,1);(1,1)}=1$, i.e., the tensor product of the ground state of the two clock chains is also the ground state of the orbifold model. Moreover, for those $N_{(a,c);(b,d)}>0$ the scaling dimension of the holomorphic primary operator $(a,c)$ is $ h_{(a,c)}=h_a+h_c$ and that of the antiholomorphic primary operator $(b,d)$ is  $\bar{h}_{(b,d)}=\bar{h}_b+\bar{h}_d$. The fact that the ground state of the orbifold model remain the same as the tensor product of the ground states of the decoupled clock chains implies \be c_{\rm orbifold}=c_{\rm decoupled~clock~chains}.\ee The latter identity can be seen from the fact that the central charge can be computed from the entanglement entropy, which is a pure ground state property. Thus, after the orbifold, the system is still conformal invariant (i.e. quantum critical) and the central charge is unaffected by the orbifold. This argument allows us to conclude that the central charge of the $Z_n\times Z_n$ ($n=2,3,4$) SPT phase transition is indeed given in table I. 
\\

In appendices \ref{orbifoldz2xz2},\ref{orbifoldz3xz3} and \ref{orbifoldz4xz4}  we go through the details of obtaining the modular invariant partition function for the orbifold $Z_n\times Z_n$ ($n=2,3,4$) clock chains. We examine the primary scaling operator content of the modular invariant conformal field theory. In addition, we study the symmetry transformation properties of various Verma modules and the scaling dimension of primary scaling operators, in particular, that of the gap opening operator. In Table \ref{tab:alllevels} we list the first few most relevant scaling operators and their scaling dimension for $n=2,3,4$. Entries in blue are invariant under $Z_n\times Z_n$.

\begin{table*}[h]
\centering
\caption{(Color online) The first few primary operators, with the lowest scaling dimensions ($h+\bar{h}$), of the orbifold $Z_n\times Z_n$ CFT for $n=2,3,4$. The  momentum quantum numbers of these operators are equal to $(h-\bar{h})\times 2\pi/N$. Entries in blue are invariant under $Z_n\times Z_n$.}
\vspace{0.1 in}
\begin{tabular}{|c|c|c|c|}
\hline
$n$					&$h+\bar{h}$		&$h-\bar{h}$	&Multiplicity \rule{0pt}{2.5ex}	\\
\hline
\multirow{5}{*}{2}	&\B{0}				&0				&\B{1}\\
\cline{2-4}
					&1/4				&0				&2\\
\cline{2-4}
					&\B{1}				&0				&2+\B{2}\\
\cline{2-4}
					&1					&$\pm1$			&2\\
\cline{2-4}
					&5/4				&$\pm1$			&8\\
\hline
\hline

\multirow{9}{*}{3}		&\B{0}				&0				&\B{1}\\
\cline{2-4}
		&4/15				&0				&4\\
\cline{2-4}
		&\B{4/5}			&0				&\B{2}\\
\cline{2-4}
		&14/15				&0				&4\\
\cline{2-4}
		&17/15				&$\pm 1$		&8\\
\cline{2-4}
		&19/15				&$\pm1$			&16\\
\cline{2-4}
		&4/3				&0				&4\\
\cline{2-4}
		&22/15				&0				&8\\
\cline{2-4}
		&\B{8/5}			&0				&\B{1}\\
\hline
\hline
\multirow{12}{*}{4}		&\B{0}				&0				&\B{1}\\
\cline{2-4}
		&1/4				&0				&4\\
\cline{2-4}
		&1/2				&0				&2\\
\cline{2-4}
		&5/8				&0				&4\\
\cline{2-4}
		&\B{1}				&0				&\B{2}\\
\cline{2-4}
		&9/8				&$\pm1$			&8\\
\cline{2-4}
		&5/4				&0				&20\\
\cline{2-4}
		&5/4				&$\pm1$			&12\\
\cline{2-4}
		&5/4				&$\pm1$			&16\\
\cline{2-4}
		&3/2				&$\pm1$			&8\\
\cline{2-4}
		&13/8				&0				&4\\
\cline{2-4}
		&13/8				&$\pm1$			&16\\
\hline
\end{tabular}
\label{tab:alllevels}
\end{table*}

\section{Numerical DMRG study of the $Z_3\times Z_3$ SPT phase transition\\}

In this section, we report  the results of numerical density matrix renormalization group calculation for the $Z_3\times Z_3$ transition. The purpose is to check our analytic predictions in the last section. The details of the numerical calculations are presented in appendix \ref{DMRG}.\\

\begin{figure}[h!]
\centering
\includegraphics[scale=0.30]{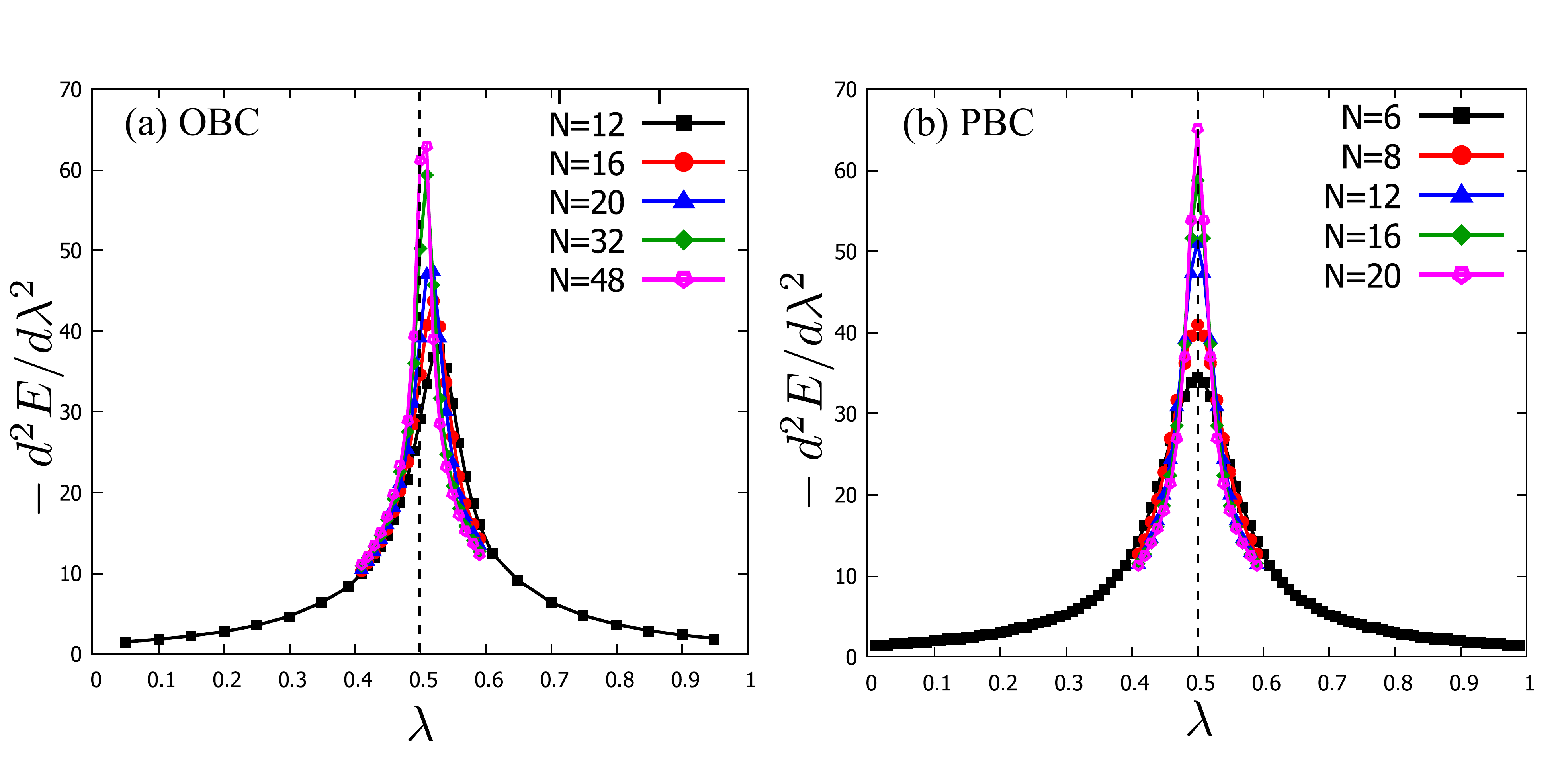}
\caption{(Color online) Second order derivative of the ground state energy with respect to $\lambda$ for both open (OBC) and periodic (PBC) boundary conditions and different values of $N$. The results suggest a divergent $-\dd[2]{E}{\lambda}$ as $\lambda\ra 1/2$ and $N\ra\infty$. Hence it signifies a second-order phase transition. As expected, we note that finite size effect is significantly stronger for open as compared to periodic boundary condition. }
\label{fig:z3phase}
\end{figure}

First, we demonstrate that $\lambda=1/2$ in \Eq{eq:hlambda} is indeed a critical point. Let's look at the second derivative of the ground state energy with respect to $\lambda$ for both open and periodic boundary conditions with different system sizes (\Fig{fig:z3phase}). The results clearly suggest a second-order phase transition at $\lambda_c=1/2$ where the second order energy derivative diverges. \\

\begin{figure}[h!]
\centering
\includegraphics[scale=0.30]{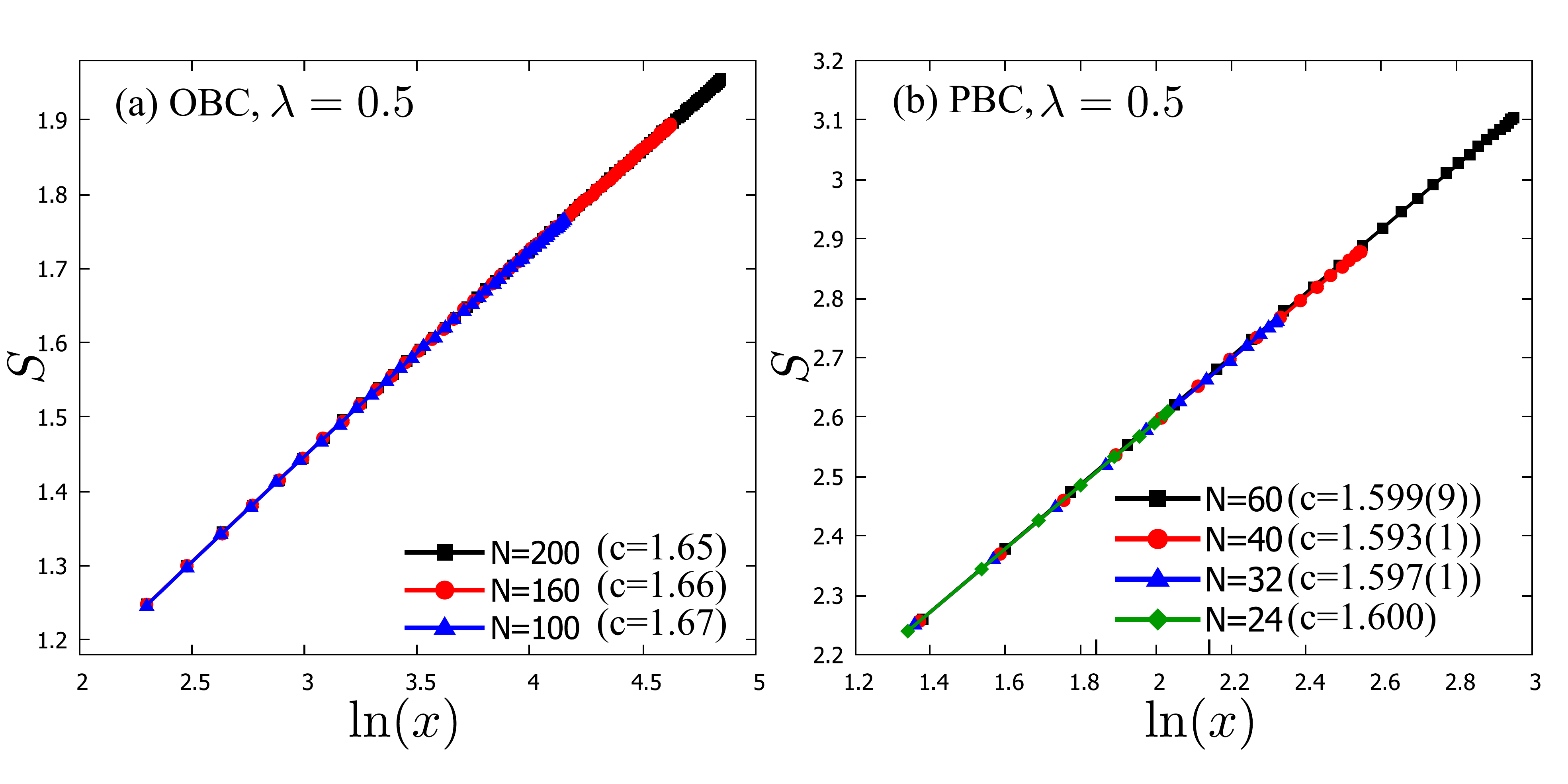}
\caption{(Color online) Entanglement entropy is plotted against $\ln(x)$, (where $x=\frac{N}{\pi}\sin(\pi l /N)$ and $l$ is the size of the subsystem which is not traced over) for a few different total system length $N$. (a) For  open boundary condition (OBC) the maximum $N$ is $200$. (b) For periodic boundary condition (PBC) the maximum $N$ is $60$. Combining these results we estimate $c=1.62\pm 0.03$.}
\label{fig:z3c}
\end{figure}

Next, we compute the central charge at $\lambda=1/2$. This is done by computing the entanglement entropy, which is calculated from the reduced density matrix by tracing out the degrees of freedom associated with $N-l$ sites in a system with total $N$ sites. In \Fig{fig:z3c} we plot the von Neumann entanglement entropy $S$ against $x=\frac{N}{\pi}\sin(\pi l /N)$ where $l$ is the number of sites that are not traced out. CFT predicts $S=\frac{c}{6}\ln(x)+const$ for the open boundary condition and $S=\frac{c}{3}\ln(x)+const$ for periodic boundary condition\cite{CFTCC2004}. From the numerics we find $c=1.599(9)$. This result is in nearly perfect agreement with our analytic prediction $c=8/5$.\\ 

\begin{figure}[h!]
\centering
\includegraphics[scale=0.30]{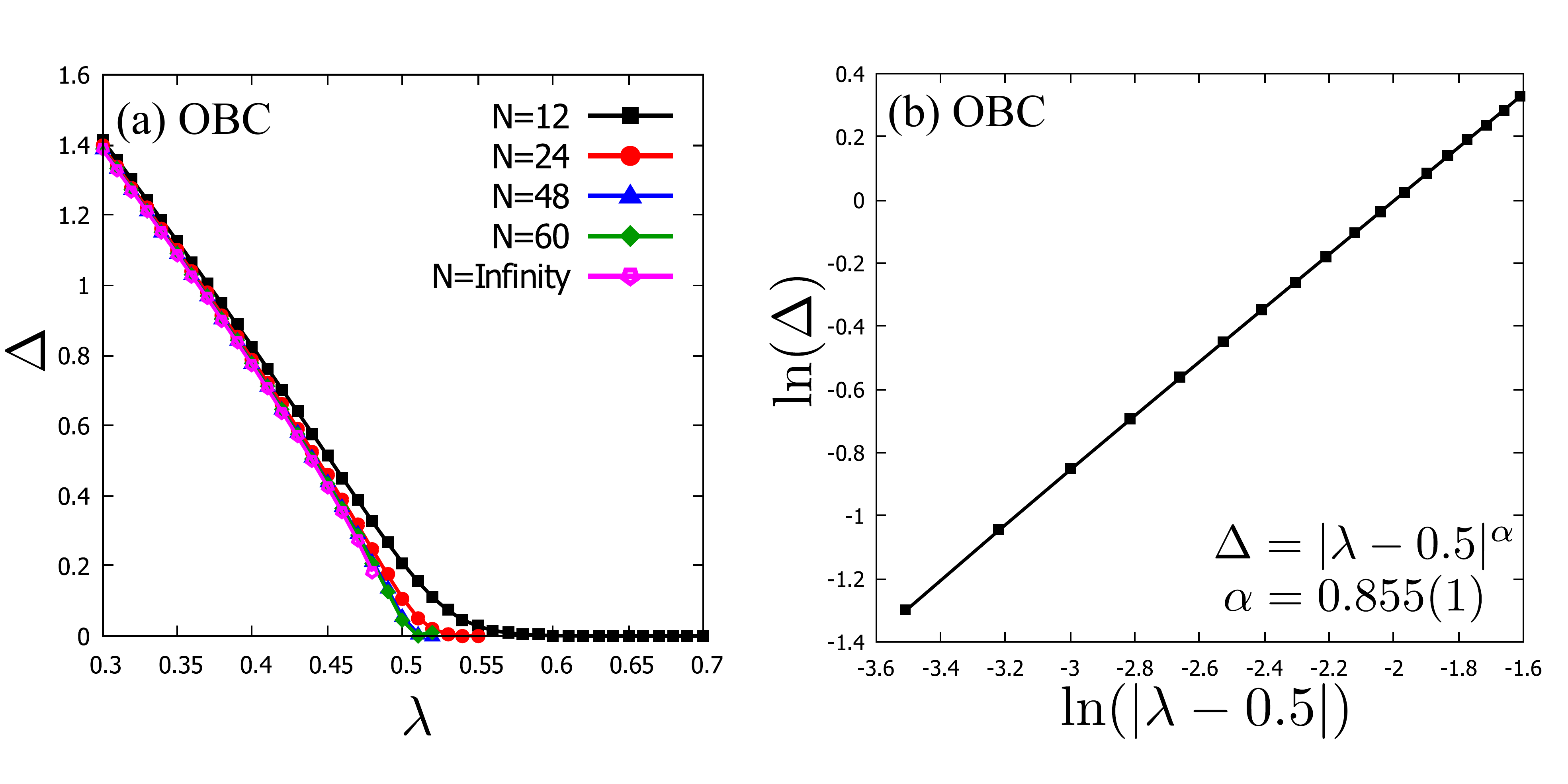}
\caption{(Color online) The energy gap $\Delta$ as a function of $\lambda$ for open boundary condition. (a) The gap closes for $\lambda>1/2$ because of the presence of edge modes associated with the non-trivial SPT. (b) The gap exponent is extracted by approaching $\lambda_c$ from the $\lambda<1/2$ side. The value of $\alpha$ is found to be 0.855(1).}
\label{fig:z3gapobc}
\end{figure}

In addition to the above results, we have also calculated the gap as a function of $\lambda$. In fitting the result to \be \Delta\sim |\lambda-\lambda_c|^\alpha\ee we estimate the gap exponent to be $\alpha=0.855(1)$ for open boundary condition (\Fig{fig:z3gapobc}) and $\alpha=0.847(1)$ for periodic boundary condition (\Fig{fig:z3gappbc}). These results are  in good agreement with the analytic prediction $\alpha=5/6$ (see appendix \ref{appdx:scaldim}).

\begin{figure}[h!]
\centering
\includegraphics[scale=0.30]{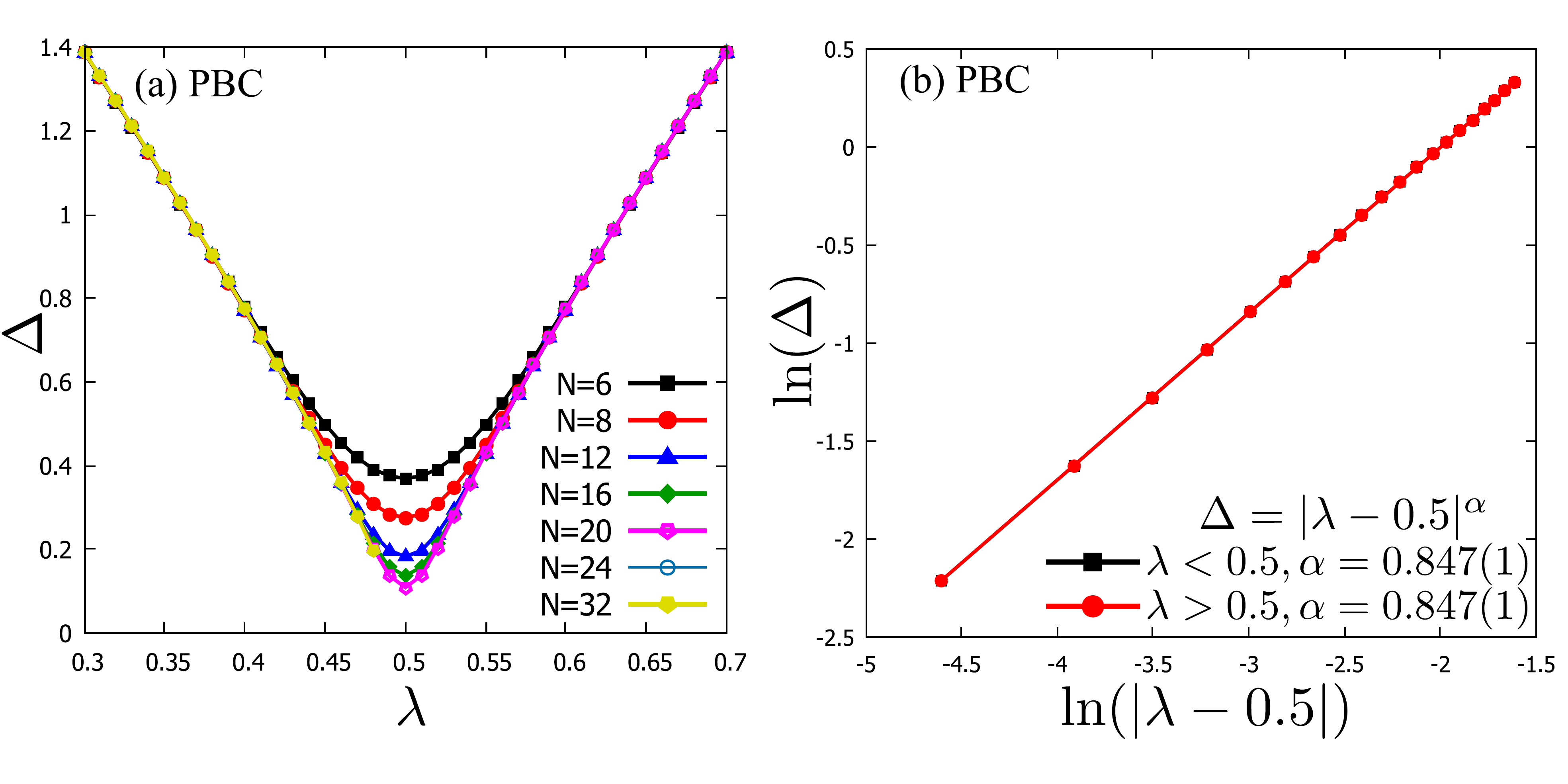}
\caption{(Color online) The energy gap $\Delta$ as a function of $\lambda$ for periodic boundary condition. (a) Now there is a non-zero gap for both $\lambda>1/2$ and $\lambda<1/2$. (b) The gap exponent is extracted and found to be $\alpha=0.847(1)$.}
\label{fig:z3gappbc}
\end{figure}

\section{The constraint on the central charge}
After an examination of Table I it is easy to notice that $c\ge 1$ for all $Z_n\times Z_n$ SPT phase transitions. Moreover, for all the cases we know, including SPTs protected by  continuous groups, all 1D ($z=1$) bosonic SPT phase transitions are described by CFT with $c\ge 1$. In the following present an argument that {\it the CFT of all 1D bosonic SPT phase transition must have $c\ge 1$.}  \\

We proceed by showing that the $c< 1$ CFTs cannot be the critical theory for bosonic SPT transitions. The 1D CFTs that are {\it unitary} and have $c<1$ are the so-called minimal models. In appendix \ref{appdx:symm} we summarize the argument in Ref.\onlinecite{Ruelle1998} where it is shown that the {\it maximum} on-site internal symmetry (``on-site'' symmetries are the ones consisting of product over local transformations that act on the local, e.g. site or group of sites, Hilbert space.) that these CFTs can possess are either $Z_2$ or $S_3$. Since the critical point of the bosonic SPT phase transitions must possess the same on-site symmetry as the phases on either side, and neither $Z_2$ nor $S_3$ can protect non-trivial bosonic SPTs in 1D (i.e., $H^2(Z_2,U(1))=H^2(S_3,U(1))=Z_1$), we conclude that the CFTs corresponding to the minimal model cannot possibly be the critical theory for bosonic SPT phase transitions. This leaves the $c\ge 1$ CFTs the only possible candidates as  the critical theory for bosonic SPT phase transitions.

\section{Conclusions}

In this paper, we present an analytic theory for the phase transition between symmetry protected topological states protected by the $Z_n\times Z_n$ symmetry group. We have shown that for $2\le n\le 4$ a direct, continuous, topological  phase transition exists. In contrast for $n\ge 5$ the transition from the topological trivial to non-trivial SPTs is intervened by an intermediate gapless phase. Our theory predicts that for $n=2,3,4$ the central charge of the CFT describing the SPT phase transitions are $c=1,8/5$ and $2$, respectively. We perform explicit numerical density matrix renormalization group calculations for the interesting case of $n=3$ to confirm our analytic predictions.\\ 

We expect treatment analogous to what's outlined in this paper can be generalized to the phase transitions between SPTs protected by symmetry group $Z_{n_1}\times Z_{n_2}\times ...$. In addition, we provide the proof for a conjectured put forward in a previous unpublished preprint\cite{Tsui2015} that the central charge of the CFTs describing bosonic SPT transitions must be greater or equal to 1. Thus all $c<1$ CFTs cannot be the critical theory for bosonic phase transitions. However, we have not yet answered the question 
``are all CFTs with $c\ge 1$ capable of describing topological phase transitions.''\\

Of course upon non-local transformation the $c<1$ minimal models can be viewed as the critical theory for {\it parafermion} SPT transitions. Indeed, the $c=1/2$ Ising conformal field theory describes the critical Majorana chain. The $c=4/5$ three-state Potts model CFT describes the critical point of $Z_3$ parafermion chain. We suspect that the parafermion models escape the classification of either the K theory or the cohomology group because of its non-local commutation relation. \\  

In space dimension greater than one, we do not know a model which {\it definitively} exhibits a continuous phase transition between bosonic SPTs. This is due partly to the likelihood of spontaneous breaking of the discrete protection symmetry when $d\ge 2$. In addition, even if the continuous phase transition exists, it is more difficult to study these phase transitions, even numerically. However a ``holograpic theory'' was developed for phase transitions between SPT phases which satisfy the ``no double-stacking constraint''\cite{Lokman2015}. That theory predicts the critical point should exhibit ``delocalized boundary excitations'' of the non-trivial SPT, which are extended ``string'' or ``membrane'' like objects with gapless excitation residing on them. We expect this kind of critical point to be fundamentally different from the Landau-like critical point. Clearly many future studies are warranted for the understanding of these interesting phase transitions.

\appendix
\newpage

\section{Cocycles, projective representation and the construction of fixed point $Z_n \times Z_n$ SPT Hamiltonian in 1D}\label{appdx:znznh}

We briefly review the definition of cocycles in the group cohomology, and describe a procedure\cite{Chen2011,Ran2012} to construct fixed point SPT hamiltonians \eqref{eq:znznh} that are relevant to this paper.\\

\subsection{Cocycle}

In 1D a cocycle associated with group $G$ is an $U(1)$ valued function $ \nu(g_0,g_1,g_2)$ where the argument $g_i\in G$ which satisfies $\nu(gg_0,gg_1,gg_2)=\nu(g_0,g_1,g_2) $. Here we only consider the group realised by unitary representation. Moreover, $\nu$ satisfies the following cocycle condition
\be(\p\nu)(g_0,g_1,g_2,g_3):={\nu(g_1,g_2,g_3)\nu(g_0,g_1,g_3)\over \nu(g_0,g_2,g_3)\nu(g_0,g_1,g_2)}
=1.\label{cocycle} \ee If \be\nu(g_0,g_1,g_2)=\p c(g_0,g_1,g_2):=c(g_1,g_2)c(g_0,g_1)/c(g_0,g_2)\label{cb}\ee for certain 
$c(g_0,g_1)$ satisfying $c(gg_0,gg_1)=c(g_0,g_1)$  we say it is a coboundary. It may be checked that a coboundary automatically satisfies the cocycle condition \Eq{cocycle}. Two cocycles related by the {\it multiplication} of a coboundary are viewed as equivalent.
\be
\nu\sim \nu^\prime~~{\rm if~~} \nu^\prime= \nu\cdot \p c.
\ee
The equivalence classes of cocycles form $H^{2}(G,U(1))$ -- the $2^{\rm nd}$ cohomology group of $G$ with $U(1)$ coefficient. Bosonic $G$-symmetric SPTs in $1$ space dimensions are ``classified'' by $H^{2}(G,U(1))$, i.e., each equivalent class of SPTs is in one to one correspondence with an element of the abelian group $H^{2}(G,U(1))$. The binary operation of the abelian group corresponds to the ``stacking'' operation, i.e., laying two SPTs on top of each other and turning on all symmetry allowed interactions.\\

\subsection{Projective representation}\label{appdx:projrep}
In quantum mechanics, symmetry operators are usually realised as matrices  $\mathcal{R}(g)$ acting on Hilbert space. Usually these matrices form a {\it linear} representation of the symmetry group, namely, \be\mathcal{R}(g_1)\mathcal{R}(g_2)=\mathcal{R}(g_1g_2).\label{lin}\ee 
However, two quantum states differ by an $U(1)$ phase are regarded as the same quantum mechanically. Thus, one should relax \Eq{lin} by allowing a phase ambiguity $\w$, namely,
\be
\mathcal{R}(g_1)\mathcal{R}(g_2)= \omega(g_1,g_2) \mathcal{R}(g_1g_2).\label{prj}
\ee
When \Eq{prj} is satisfied we say that $\mathcal{R}(g)$ form a {\it projective} representation of the 
original symmetry group.  Obviously, linear representation where $\w(g_1,g_2)=1$ is a special case of projective representation. In the literature linear representations are usually viewed as ``trivial'' projective representations. Associativity under group multiplication, namely,
\be
[\mathcal{R}(g_1)\mathcal{R}(g_2)]\mathcal{R}(g_3)=\mathcal{R}(g_1)[\mathcal{R}(g_2) \mathcal{R}(g_3)]
\ee
requires
\be
\omega(g_1,g_2g_3)\omega(g_2,g_3)=&\omega(g_1,g_2) \omega(g_1g_2,g_3)
\label{assc}
\ee
In addition the phase ambiguity of quantum states obviously allows one to multiply all $\mathcal{R}(g)$ by an $U(1)$ phase $\phi(g)$,
namely,
\begin{align*}
\mathcal{R}(g) \rightarrow & \phi(g) \mathcal{R}(g).
\end{align*}
This phase transformation results in 
\be
\omega(g_1,g_2) \rightarrow &  \frac{\phi(g_2) \phi(g_1)}{\phi(g_1g_2)}\omega(g_1,g_2)
\label{equivc}
\ee
\noindent Consequently $\omega$s related by \Eq{equivc} 
should be regarded as equivalent. 
\\

It turns out that in 1D, cocycles of group cohomology can be interpreted as projective representations. The easiest way to see it is by defining $\omega(g_1,g_2)$ and $\phi(g_1)$ in terms of the cocycle $\nu$ and the coboundary $c$ defined in the last subsection, namely, 

\begin{align*}
\omega(g_1,g_2):=&\nu (e,g_1,g_1g_2) \\
\phi(g_1):=&c (e,g_1). 
\end{align*}

\noindent where $e$ is the identity group element of $G$. In terms of $\w$ the cocycle condition becomes

\be
&&(\p \omega)(g_1,g_2,g_3):=(\p \nu) (e,g_1,g_1 g_2,g_1g_2g_3)\nn 
&&= \frac{\nu (g_1,g_1g_2,g_1g_2g_3)}{\nu (e,g_1g_2,g_1g_2g_3)} \frac{\nu (e,g_1,g_1g_2g_3)}{\nu (e,g_1,g_1g_2)}\nn
&&=\frac{\omega(g_2,g_3)}{\omega(g_1g_2,g_3)} \frac{\omega(g_1,g_2g_3)}{\omega(g_1,g_2)}=1\nn
&&\implies \omega(g_1,g_2g_3)\omega(g_2,g_3)=\omega(g_1,g_2) \omega(g_1g_2,g_3),
\ee
namely \Eq{assc}.
In terms of $c$ the coboundary equivalence relation becomes
\be
&&\omega \sim \omega^\prime~~{\rm if~~} \omega^\prime= \omega \cdot \p \phi,\ee where
\be
&&(\p \phi)(g_1,g_2):=(\p c)(e,g_1,g_1g_2) \nn
&&=\frac{c(g_1,g_1g_2) c(e,g_1)}{c(e,g_1g_2)} \nn
&&=\frac{\phi(g_2) \phi(g_1)}{\phi(g_1g_2)},
\ee
which is exactly the factor appearing in \Eq{equivc}.

\subsection{Construction of Hamiltonian}

\begin{figure}[h!]
\centering
\includegraphics[scale=0.4]{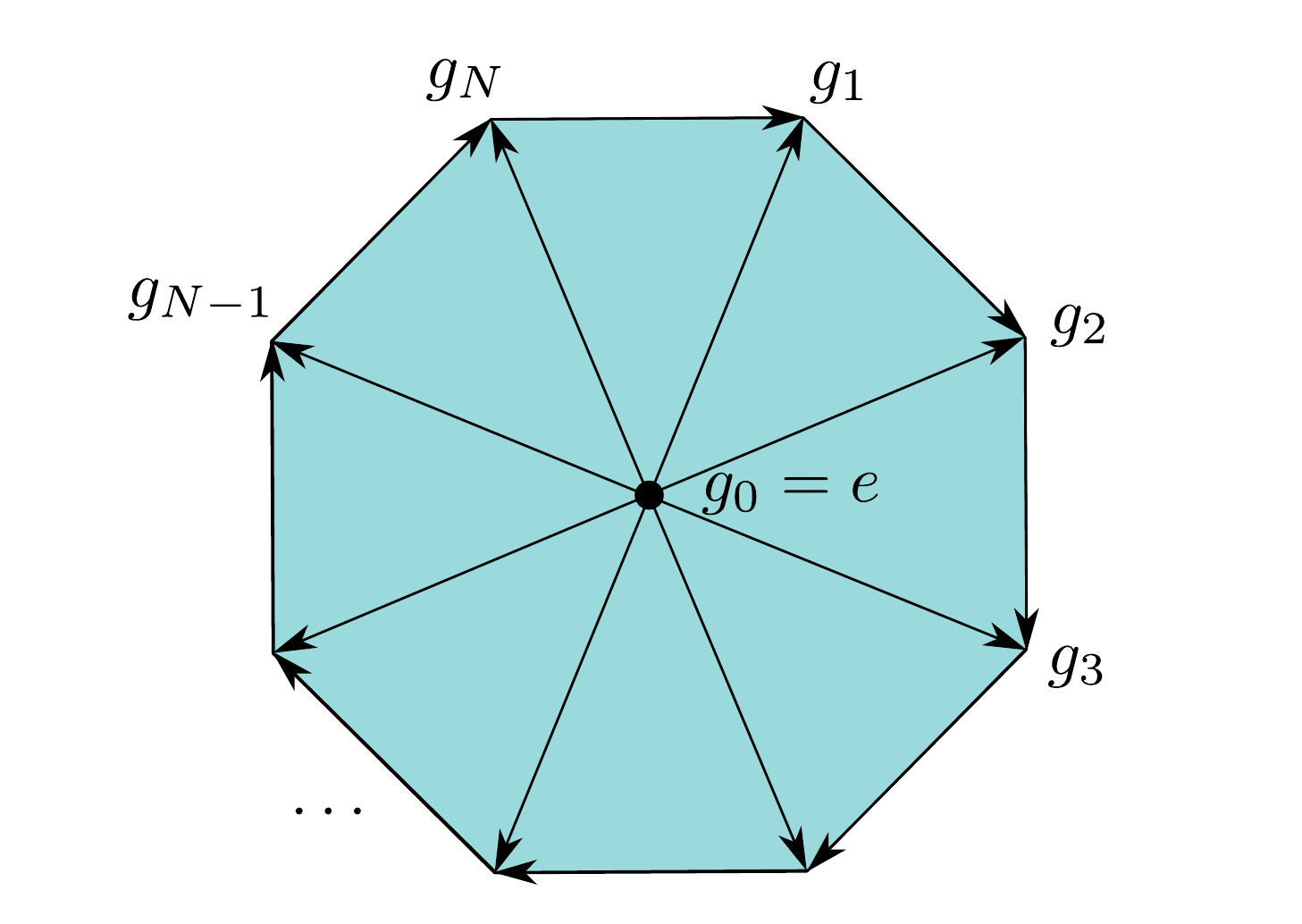}
\caption{(Color online) Construction of the 1D groundstate SPT wavefunction from the cocycle. Here the physical degrees of freedom labelled by $g_1,\dots,g_{N}$ live on the boundary of the figure. At the center, there is an auxiliary ``0" site to which we attach the identity group element $e$. A phase can be assigned to each triangle by evaluating the cocycle on the group elements on the vertices. The wavefunction is the product of the phases from all triangles.}
\label{fig:sptgs}
\end{figure}

Here we describe how to construct solvable Hamiltonians, one for each equivalence class of the SPTs. Consider a 1D ring consists of $N$ lattice sites. The Hilbert space for each site $i$ is spanned by $ \{|g_i\>\} $ where $g_i\in G$, and the total Hilbert space is spanned by the tensor product of the site basis, i.e., $|\{g_i\}\>=\prod_i|g_i\>$. For each class of the SPTs (or for each element of  $H^{2}(G,U(1))$) picks a representing cocycle $\nu(g_0,g_1,g_2)$. The ``fixed point'' ground state, which is a particular representative of a whole equivalent class of SPTs, associated with the cocycle $\nu$ is equal to (\Ref{Chen2011} Section IX)
 \be
&&|\psi_0\rangle=\sum_{\{g_i\}}\phi(\{g_i\})~|\{g_i\}\rangle, {\rm ~~where}\nn
&&\phi(\{g_i\})=\prod_{i=1}^L[\nu(e,g_i,g_{i+1})]^{\sigma(0,i,i+1)}.
\label{eq:gswf}
\ee
Here $ e $ represents the identity element of $G$. It is attached to ``0'' site at the center of the ring as shown in \Fig{fig:sptgs}.  $ \sigma(i,i+1) =\pm 1$ depending on the orientation of the triangle $0,i,i+1$. The orientation of each link in the triangle is represented by an arrow pointing from the site labeled by a smaller site index to the site labeled by the bigger index. From the link orientation we determine the triangle orientation by following the majority of the link orientation and the right-hand rule). Finally periodic boundary condition requires $g_{N+1}=g_1$. \\

The Hamiltonian whose exact ground state is \Eq{eq:gswf} is
\begin{align}
H=-J \sum_i B_i ,\label{eq:bulkh}
\end{align}
where $J>0$. The operator $ B_i $ only changes the basis states on site $i$,  and
\be
\<\{g_k'\}|B_i|\{g_k\}\>=\left(\prod_{k\ne i}\delta_{g_k',g_k}\right)\frac{\nu(g_{i-1},g_i,g_i')}{\nu(g_i,g_i',g_{i+1})}.
\label{eq:me}
\ee

For $G=Z_n\times Z_n$, there are $n$ inequivalent SPT classes and  $H^2(Z_n\times Z_n,U(1))=Z_n$. Each equivalent class of $H^2(Z_n\times Z_n,U(1))$ is represented by a cocycle  $$\nu((e,e),(g_1,g_2),(g_3,g_4))=\eta_n^{k g_2 g_3},{\rm~~where~~}\eta_n=e^{i2\pi/n}$$ In the above  $(g_{2i-1},g_{2i})\in Z_n\times Z_n$ are the $Z_n$ elements associated with site $i$, and $k\in\{0,1,\dots,n-1\}$ each correspond to a different element of $Z_n$ ($H^2(Z_n\times Z_n,U(1))$). In the main text, we refer to $|g_{2i-1},g_{2i}\>$ as the ``cell basis'' which is the tensor product of the ``site basis'' $|g_{2i-1}\>$ and $|g_{2i}\>$. The fixed point Hamiltonian constructed using the procedure discussed above is 
\begin{align*}
H=-\sum_{i}(B_{2i-1} +  B_{2i} + h.c.),
\end{align*}
where $B_{2i-1}$ changes the state  $|g_{2i-1}\>$ and $B_{2i}$ changes the state $|g_{2i}\>$. Explicitly calculating the matrix element (\Eq{eq:me}) for the cases, i.e.,  $k=0$ and $k=1$, relevant to our consideration (recall that we are interested in the quantum phase transition between SPTs correspond to the ``0'' and ``1'' elements of $Z_n$) it can be shown that 
\be
&&k=0:~~B_{2i-1}=M_{2i-1},~~~~~~~~~~~~~~B_{2i}=M_{2i}\nn
&&k=1:~~B_{2i-1}=R^{\dagger}_{2i-2} M_{2i-1} R_{2i},~~B_{2i}=R_{2i-1} M_{2i} R^{\dg}_{2i+1},
\ee
where $M_j$ and $R_j$ are defined by \Eq{mr} of the main text.

\section{The mapping to $Z_n\times Z_n$ clock models with spatially twisted boundary condition and a Hilbert space constraint}\label{appdx:of}

In this section, we show that \Eq{eq:hdimer} and \Eq{eq:hlambda} of the main text can be mapped onto an
``orbifold'' $Z_n\times Z_n$ clock models. The mapping is similar to the ``duality transformation'' of the $Z_n$ clock model. The mapping is given by 
\be
&&R^{\dagger}_{j-1}R_{j}=\widetilde{M}_j,~{\rm for}~~j=2 ...2N,~{\rm and}~R^{\dagger}_{2N}R_{1}= \widetilde{M}_{1}.
\nn 
&&M_{j}= \widetilde{R}^{\dagger}_{j} \widetilde{R}_{j+1}~~\text{for all}~~j.
\label{dual2}
\ee
Here the tilde operators obey the same commutation relation as the un-tilde ones.
Due to the periodic boundary condition on $R$, namely, $R_{2N+1}=R_1$ and $R_{2N+2}=R_2$ the line of \Eq{dual2}implies
\be
\prod _{i=1}^{2N} \widetilde{M}_i=1.
\label{dcon1}
\ee
Moreover, if we also impose the periodic boundary condition on  $\widetilde{R}_{j}$ a similar constraint on $M_i$, namely,
\be 
\prod _{i=1}^{2N}{M}_i=1.
\label{spur}
\ee
is obtained. Since there is no such constraint on $M_i$ in the original problem we need to impose a ``twisted'' boundary condition on $\widetilde{R}_{j}$:
\be
&&\widetilde{R}_{2N+1}=\tilde{B}\widetilde{R}_{1}\nn&&\widetilde{R}_{2N+2}=\tilde{B}\widetilde{R}_{2}
\label{bc1}\ee where $\tilde{B}$ commutes with all $\tilde{R}_j$ and $\tilde{M}_j$. Moreover $\tilde{B}$ has eigenvalues $\tilde{b}=1,\eta_n,...,\eta_n^{n-1}$, i.e., $$\tilde{B}|\tilde{b}\>=\tilde{b}|\tilde{b}\>.$$\\ 

Substituting \Eq{dual2} into \Eq{eq:hdimer} and \Eq{eq:hlambda} of the main text we obtain the following expression of the transformed Hamiltonian
\be
&&H_{01}(\lambda)=H^{\rm even}_{01}(\lambda)+H^{\rm odd}_{01}(\lambda)\text{~~where}\nn
&&H^{\rm even}_{01}(\lambda)=-\sum\limits_{i=1}^{N}\left[(1-\lambda)\widetilde{M}_{2i}+\lambda \widetilde{R}^{\dagger}_{2i}\widetilde{R}_{2i+2} \right]+ h.c.\nn
&&H^{\rm odd}_{01}(\lambda)=-\sum\limits_{i=1}^{N}\left[\lambda\widetilde{M}_{2i+1}+(1-\lambda) \widetilde{R}^{\dagger}_{2i-1}\widetilde{R}_{2i+1} \right]+ h.c.
\label{dualH3}
\ee
It is important to note that \Eq{dualH3} is supplemented with the spatially twisted boundary condition 
\be
\widetilde{R}_{2N+1}:= \widetilde{B}\widetilde{R}_{1}\text{~~and~~} \widetilde{R}_{2N+2}:= \widetilde{B}\widetilde{R}_{2}
\label{eq:twist2}
\ee
and the constraint \Eq{dcon1}. In addition, after the transformation the two generators of the $Z_n \times Z_n$ group become \\
\begin{align}
\widetilde{B}\text{~~~and~~~}\prod_{j=1}^N \widetilde{M}_{2j}. \label{sym}
\end{align}

On the surface \Eq{dualH3} describes two decoupled $Z_n$ clock models living on even and odd sites, respectively. 
However the notion of ``decoupled chains'' is deceptive because the constraint in \Eq{dcon1} couples them together.

\section{The notion of ``orbifold''}\label{appdx:orbifold} 
A useful way to implement the constraint \Eq{dcon1} is to apply the projection operator 
\be
\frac{1}{n}\sum_{q=0}^{n-1} Q^q\label{proj}\ee to states in the Hilbert space, where the operator $Q$ is given by
\be Q:=\prod_{j=1}^{2N} \widetilde{M}_j. \label{here}\ee Because the eigenvalues of $Q$ are $1,\eta_n,...,\eta_n^{n-1}$.
\Eq{proj} projects onto those states in the Hilbert space that are symmetric under the action of $Q$. 
The partition function of the Hamiltonian \eqref{dualH3}, subject to constraint \Eq{dcon1} and summed over twisted spatial boundary condition sectors is therefore
\be
Z=\frac{1}{n}\sum_{q_\tau=0}^{n-1}\sum_{q_s=0}^{n-1}\Tr\Big[Q^{q_\tau} e^{-\beta (H^{even}+H^{odd})}\Big]_{q_s},\label{part1}\ee
where $\Tr[...]_{q_s}$ denotes the trace under the spatially twisted boundary condition \be\widetilde{R}_{2N+1}= \eta_n^{q_s}\widetilde{R}_{1}~{\rm and}~\widetilde{R}_{2N+2}=\eta_n^{q_s}\widetilde{R}_{2}.\label{deftr}\ee
Moreover in the path integral the action of $Q^{q_\tau}$ at $\tau=\beta$ and be viewed as imposing a twisted boundary 
condition in the time direction. \\

Thus \Eq{part1} can be written as 
\be Z=\frac{1}{n}\sum_{q_s,q_\tau=0}^{n-1}Z^{\rm n-clock}_{q_s,q_\tau}\times Z^{\rm n-clock}_{q_s,q_\tau}
\label{orbifoldZ}\ee
where $Z^{\rm n-clock}_{q_s,q_\tau}$ represents clock model partition function under the space and time twisted boundary condition characterized by $q_s$ and $q_\tau$. In \Eq{orbifoldZ} $Z^{\rm n-clock}_{q_s,q_\tau}$ appears twice on right-hand side because without orbifold (i.e., summing over space and time twisted boundary conditions) we have two independent $n$-state clock models. Averaging over the partition function under space and time boundary condition {twists} is the ``orbifolding''\cite{Dijkgraaf1989}. Note that here the spatial boundary condition twist is generated by one of the $Z_n$ generator, namely $\tilde{B}$, in \Eq{sym}. However, the time twist is generated by $Q=\prod_{j=1}^{2N} \widetilde{M}_j$, which is a symmetry of the $Z_n\times Z_n$ clock Hamiltonian, \Eq{dualH3}, but it is not the generator for the other $Z_n$ in \Eq{sym}.

\section{The modular invariant partition function and the primary scaling operators of the orbifold critical $Z_2\times Z_2$ clock model}\label{orbifoldz2xz2}

\subsection{Review of modular invariant partition function for the Ising model\\}\label{appdx:Ising}
The Ising model shows an order-disorder phase transition. At the critical point, the Hamiltonian is given by
\begin{align*}
H_{Ising}=-\sum_i \left[M_i+R_{i}R_{i+1}\right]
\end{align*}
where $M_i$, $R_i$ are Pauli matrices $\s^x$ and $\s^z$ respectively (we use $M,R$ rather than $\s^x,\s^z$ for the 
consistency of notation). The central charge of a single critical Ising chain is $c=\frac{1}{2}$. Its conformal field theory is the $\mathcal{M}(4,3)$ minimal model. The primary scaling operators are labeled by two pairs of indices $(r,s)$ and $(r',s')$  each label the ``holomorphic'' and the ``anti-holomorphic'' part of the operator. The ranges of these indices are given by $1\le s\le r\le 2$ and $1\le s^\prime\le r^\prime\le 2$. 
The scaling dimensions of the holomorphic and anti-holomorphic parts of these operators are given by
\be
h_{r,s}=\frac{(4r-3s)^{2}-1}{48},~~~\bar{h}_{r',s'}=\frac{(4r'-3s')^{2}-1}{48}
\label{ihs}\ee
\Eq{ihs} gives rise to three primary holomorphic (and anti-holomorphic) scaling operators with distinct scaling dimensions. The corresponding $(r,s)$ indices are $(1,1)$, $(2,1)$ and $(2,2)$. Through the operator-state correspondence, each of these primary fields and their associated ``descendants'' form the basis of a Hilbert space (the ``Verma module'') which carries an irreducible representation of the conformal group.

\begin{figure}[h!]
\centering
\includegraphics[scale=0.4]{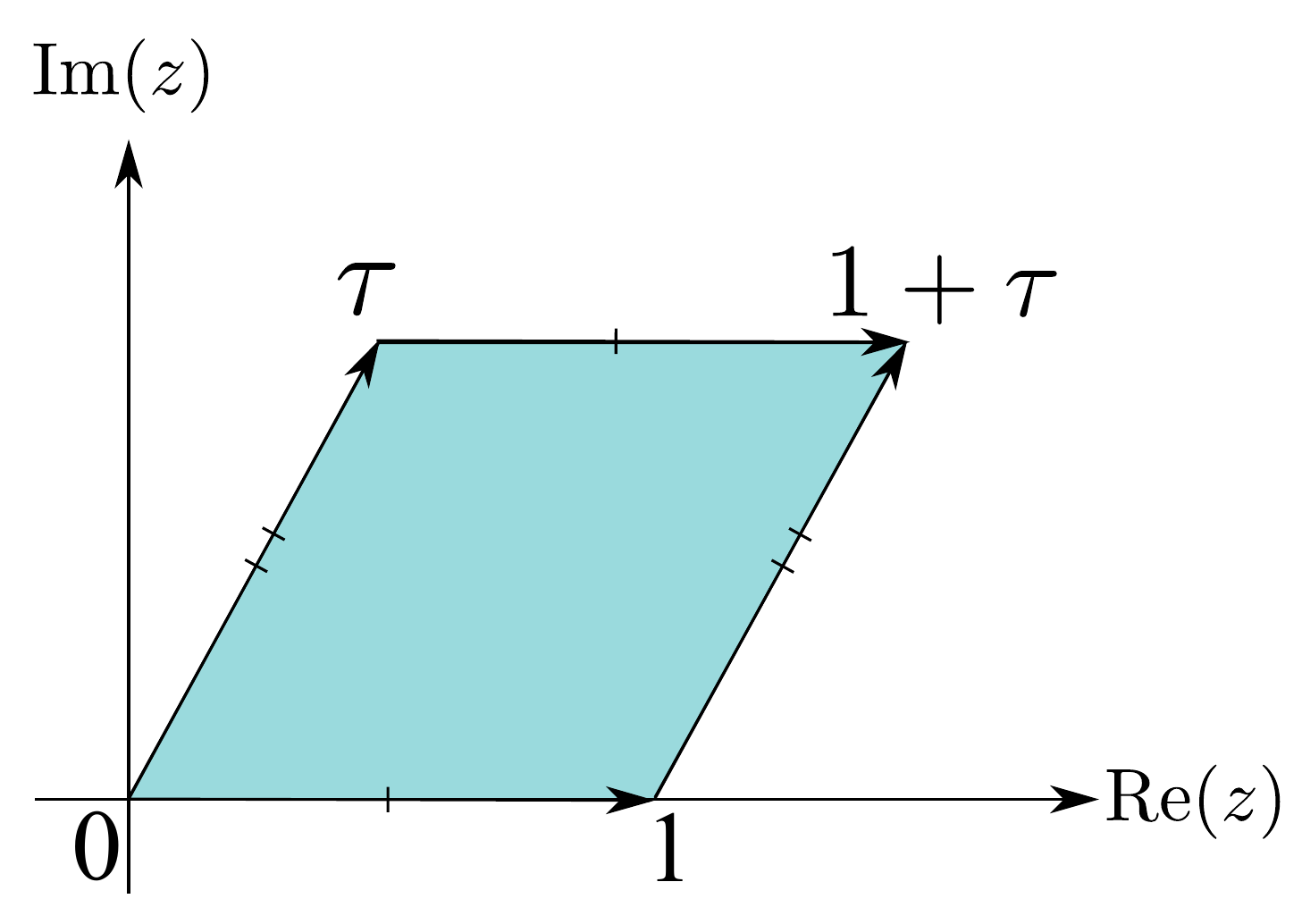}
\caption{(Color online) The spacetime torus with modular parameter $\tau$ is obtained from identifying opposite edges of a parallelogram with vertices $0,1,\tau$ and $1+\tau$ in the complex plane. Here $\tau$ is a complex number in the upper complex plane.}
\label{fig:tau}
\end{figure}

Now consider the partition function of the CFT on a spacetime torus (see \Fig{fig:tau}). The prototype torus is obtained from identifying opposite edges of the parallelogram having $(0,1,1+\tau,\tau)$  as the complex coordinates of its four vertices ($\tau\in$  upper half complex plane). On such a torus, the partition function is given by 
\be
Z(\t)=\sum_{r,s;r^\prime,s^\prime}M_{(r,s);(r^\prime,s^\prime)}\chi_{r,s}(\t)\bar{\chi}_{r^\prime,s^\prime}(\bar{\t}),\label{miz}\ee where
$M_{(r,s);(r^\prime,s^\prime)}$ is a matrix with integer entries,  and \be &&\ch_{r,s}(\t)= q^{-{c\over 24}}\Tr_{(r,s)} q^{h_{r,s}}\nn&&\bar{\ch}_{r^\prime,s^\prime}(\bar{\t})=\left(\bar{q}\right)^{-{c\over 24}}\Tr_{(r^\prime,s^\prime)} \left(\bar{q}\right)^{\bar{h}_{r^\prime,s^\prime}},\ee with $q=e^{i 2\pi\t}$ and $\bar{q}=e^{-i 2\pi\bar{\t}}$
Here the trace $\Tr_{(r,s)}$ and $\Tr_{(r^\prime,s^\prime)}$ are taken within the Verma module labeled by $(r,s)$ and $(r^\prime,s^\prime)$. In the literature $\ch_{r,s}$ and $\bar{\ch}_{r^\prime,s^\prime}$ are referred to as ``characters''. \\

For the CFT to be consistent, its partition function must be ``modular invariant''\cite{Cardy1986}. The modular group consists of discrete coordinate transformations that leave the lattice whose fundamental domain is given by {\Fig{fig:tau}} invariant. This group is generated by the $T$ ($\tau\ra\tau+1$) and the $S$ ($\tau\ra -1/\tau$) transformations. When acted upon by these transformations the characters $\chi_{{r,s}}$ (with a similar expression for $\bar{\ch}_{r^\prime,s^\prime}$) transform according to
\begin{align*}
T:~~\ch_{{r,s}}(\t+1)=\sum_{(\rho,\s)} {\rm T}_{(r,s),(\rho,\s)}~ \ch_{{\rho,\s}}(\t)\\
S:~~\ch_{{r,s}}(-1/\t)=\sum_{(\rho,\s)} {\rm S}_{(r,s),(\rho,\s)}~ \ch_{{\rho,\s}}(\t).
\end{align*}
Here  S,T are known matrices and the transformation matrices for the anti-holomorphic $\bar{\chi}$ are the complex conjugate of those of the holomorphic ones.)  \\

The requirement of modular invariance,namely,
\begin{align}
Z(\t+1)=Z(-1/\t)=Z(\t) \label{eq:modinv}
\end{align}
impose stringent constraints on the possible $M_{(r,s);(r^\prime,s^\prime)}$ in \Eq{miz}. For $c=1/2$ if we require $M_{(1,1),(1,1)}=1$, i.e., a unique ground state, there is only one such possible $M$, namely, $M_{(r,s);(r^\prime,s^\prime)}=\delta_{(r,s),(r^\prime,s^\prime)}$. The corresponding partition function is given by:
\begin{align*}
Z^{\rm Ising}(\t)=|\chi_{I}(\t)|^2+|\chi_{\e}(\t)|^2+|\chi_{\s}(\t)|^2
\end{align*}
where 
\begin{align}
&\ch_{I}:=\ch_{1,1}, ~\ch_{\e}:=\ch_{2,1}, ~\ch_{\s}:=\ch_{2,2}
\end{align}
The explicit form of $\ch_{(r,s)}$ is given by equation (8.15) of Ref.[\onlinecite{Francesco2012}].
The conformal dimensions of primary fields and their eigenvalues under the action of the $Z_2$ generator are summarized in table \ref{tab:isingsh}\cite{Francesco2012}.

\begin{table*}[h!]
\centering
\caption{Conformal dimensions of the primary fields of the Ising model, and their transformation properties upon the action of the  $Z_2$ generator.}
\vspace{0.1in}
\begin{tabular}{|c|c|c|c|}
\hline
$(r,s)$		&(1,1)		&(2,1)		&(2,2) \\
\hline
$h_{(r,s)}$	&0			&1/2		&1/16	\\
\hline
$Z_2$		&1			&1			&$-1$		\\
\hline
\end{tabular}
\label{tab:isingsh}
\end{table*}

\subsection{Constructing the orbifold partition function for the $Z_2\times Z_2$ critical theory}\label{appdx:z2orbifold}

With the brief review of the modular invariant partition function of the critical Ising model we are ready to 
construct the partition function for the orbifolded $Z_2\times Z_2$ model defined by of \Eq{orbifoldZ}: 
\be
Z_{Z_2\times Z_2}(\t)
=\frac{1}{2} \sum_{q_s=0}^1\sum_{q_\tau=0}^{1} Z^{\rm Ising}_{q_s,q_\tau}(\t)\times Z^{\rm Ising}_{q_s,q_\tau}(\t).
\label{obfI}\ee

\begin{figure}[h!]
\centering
\includegraphics[scale=0.45]{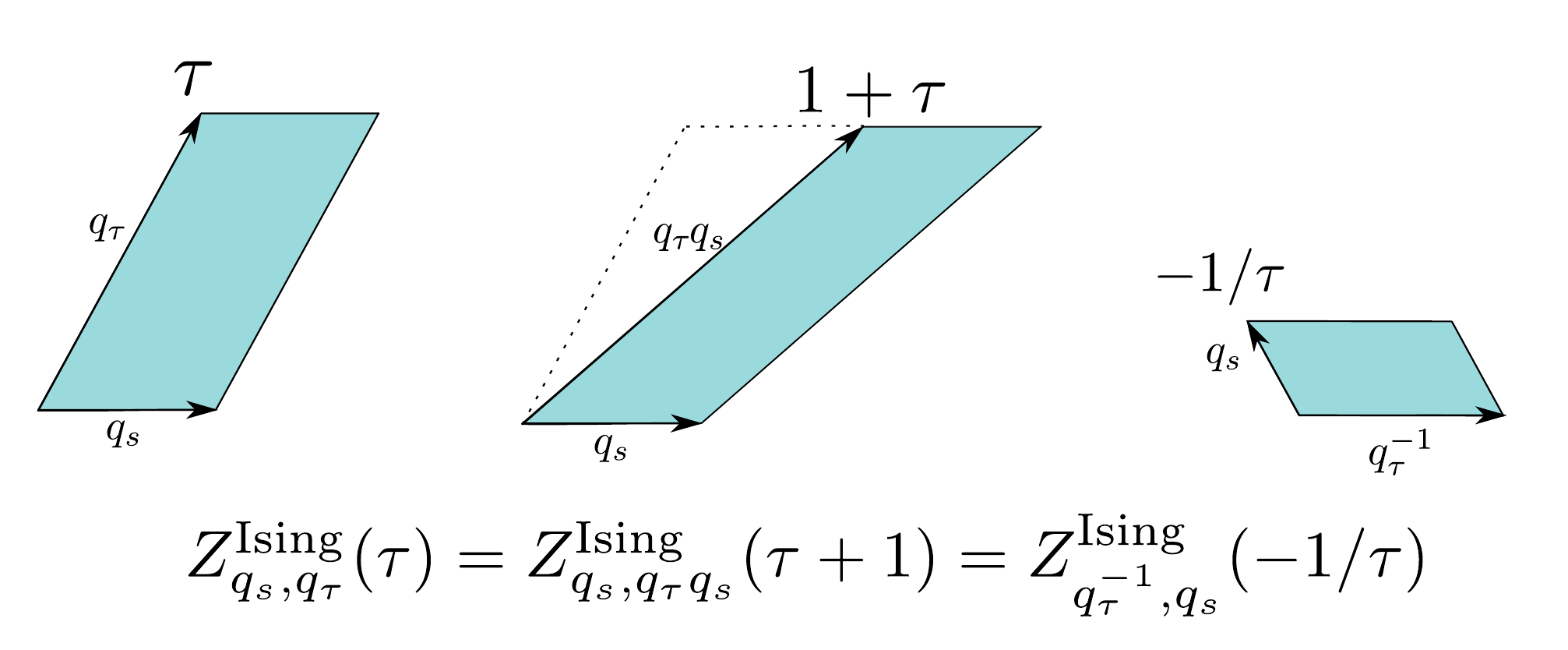}
\caption{(Color online) The transformation of the boundary twisted partition function $Z_{q_s, q_{\t}}(\t)$ under the $S$ and $T$ transformations.}
\label{fig:STtau}
\end{figure}

$Z^{\rm Ising}_{(0,1)}$ is given in Ref.[\onlinecite{Francesco2012}]. It is also shown in the same reference that $Z^{\rm Ising}_{q_s,q_{\tau}}(\t)=Z^{\rm Ising}_{q_s,q_{\tau}q_s}(\t+1)=Z^{\rm Ising}_{q_{\tau}^{-1},q_s}(-1/\t)$ (see \Fig{fig:STtau}), hence
\be
&&Z^{\rm Ising}_{0,0}(\t)=|\chi_{I}(\t)|^2+|\chi_{\e}(\t)|^2+|\chi_{\s}(\t)|^2\nn
&&Z^{\rm Ising}_{0,1}(\t)=|\chi_{I}(\t)|^2+|\chi_{\e}(\t)|^2-|\chi_{\s}(\t)|^2\nn
&&Z^{\rm Ising}_{1,0}(\t)=\mathcal{S}Z^{\rm Ising}_{0,1}(\t)\nn
&&Z^{\rm Ising}_{1,1}(\t)=\mathcal{T}Z^{\rm Ising}_{1,0}(\t)
\label{zij}
\ee
Using the known S,T matrices of the Ising model we can compute $Z^{\rm Ising}_{1,0}$ and $Z^{\rm Ising}_{1,1}$. Substitute the results into \Eq{obfI} we obtain the orbifolded $Z_2\times Z_2$ partition function:
\be
Z_{Z_2\times Z_2}(\t)=(|\ch_{I}|^{2}+|\ch_{\e}|^{2})^{2}+2|\chi_{\s}^{2}|^{2}+(\bar{\ch}_{I}\ch_{\e}+\bar{\ch}_{\e}\ch_{I})^{2}
\label{z2z2z}
\ee
where the $\t$ dependence is suppressed.  When expanded in terms of $\ch_{r,s}\bar{\ch}_{r^\prime,s^\prime}$  the first term yields 4 terms (henceforth referred to as group I terms). The second term yields 2 terms (group II terms). The third term yields 4 terms (group III terms). Due to the prefactor 2 in the second term on the right-hand side of \Eq{z2z2z}, terms in group II appear with multiplicity 2. It turns out that this partition function is the same as the $XY$ model. The first few energy levels with $h+\bar{h}<2$ and their quantum numbers are listed in Table \ref{tab:z2levels}.

\begin{table*}[h]
\centering
\caption{The quantum numbers of the first few primary operators of the orbifold $Z_2\times Z_2$ CFT.}
\vspace{0.1 in}
{\begin{tabular}{|c|c|c|c|c|}
\hline
$h+\bar{h}$		&$h-\bar{h}$	 &Multiplicity				&Terms in $Z_{Z_3\times Z_3}$	\rule{0pt}{2.5ex}				\\
\hline
0								&0			&1				&$|\ch_{I}|^{4}$			 \\
\hline
1/4								&0			&2			&$2|\ch_{\s}|^{4}$			\\
\hline
1								&0			&4				&$4|\ch_{I}|^{2}|\ch_{\e}|^{2}$\\
\hline
1								&$\pm1$		&2			&$\bar{\ch}_{I}^{2}\ch_{\e}^{2}+ \text{c.c.}$	\\
\hline
5/4							&$\pm1$			&8			&$2|\ch_{\s}|^{4}$(due to the first descendant)				\\
\hline
\end{tabular}
}
\label{tab:z2levels}
\end{table*}

\subsection{Transformation properties under the action of $Z_2 \times Z_2$} \label{appdx:z2z2transform}
To see how the {contributing} Verma modules of \Eq{z2z2z} transform under the action of  $Z_2 \times Z_2$, we construct operators that project the Hilbert space into subspaces carrying various irreducible representations of $Z_2\times Z_2$. Let 
$G_A=\widetilde{B}$ and $G_B=\prod_{i \in even} \widetilde{M}_i$ be the generators of $Z_2\times Z_2$. 
The operator that  
projects into subspace with eigenvalues $(\eta_2^{a},\eta_2^{b})$ (here $\eta_2=-1$ and $a,b=0,1$) under the action of $G_A$ and $G_B$ is given by
\be
P_{ab}=\left(\frac{1+\eta_2^{-a}G_A}{2}\right) \left(\frac{1+\eta_2^{-b}G_B}{2}\right) \ee
\\

To filter out the Verma modules that transform according to this particular irreducible representation, we compute 

\begin{align}
P_{ab}Z_{Z_2\times Z_2}&:=\frac{1}{2}\sum_{q_\tau=0}^{1}\sum_{q_s=0}^{1}\Tr\Big[P_{ab}Q^{q_\tau} e^{-\beta (H^{\rm even}+H^{\rm odd})}\Big]_{q_s}\nn
&=\frac{1}{8}\sum_{\mu,\nu=0}^{1}\sum_{q_\tau=0}^{1}\sum_{q_s=0}^{1}\eta_2^{-a\mu-b\nu}\Tr\Big[G_A^{\mu}G_B^{\nu}Q^{q_\tau} e^{-\beta (H^{\rm even}+H^{\rm odd})}\Big]_{q_s}\nn
&=\frac{1}{8}\sum_{\mu,\nu=0}^{1}\sum_{q_\tau=0}^{1}\sum_{q_s=0}^{1}\eta_2^{-a\mu-b\nu}\Big[\eta_2^{q_s\mu}(Z^{\rm Ising}_{q_s,q_{\t}})(Z^{\rm Ising}_{q_s,q_{\t}+\nu})\Big]
\label{z2partf} 
\end{align}
For example, \be P_{00}Z_{Z_2\times Z_2}=(|{\ch_{I}}|^{2}+|{\ch_{\e}}|^{2})^{2},\label{sya}\ee which means only group I transform as the identity representation of $Z_2 \times Z_2$. For other $P_{ab}$ the results are summarized in table \ref{tab:z2charges}
\\

\begin{table*}[h]
\centering
\caption{Transformation properties of the contributing Verma modules in \Eq{z2z2z} under the action of $G_A$ and $G_B$. For group II, the doublet records the transformation properties of the multiplicity two Verma modules in \Eq{z2z2z} .}
\vspace{0.1 in}
\begin{tabular}{|c|c|c|}
\hline
Group	&$G_A$	&$G_B$ \\
\hline
I		&1			&1\\
\hline
II		&$(1,-1)$		&$(-1,1)$\\
\hline
III		&$-1$		&$-1$\\
\hline
\end{tabular}
\label{tab:z2charges}
\end{table*}

\subsection{Scaling Dimension for the operator driving the $Z_2\times Z_2$ SPT transition}\label{appdx:z2scaldim}
\begin{figure}[h!]
\centering
\includegraphics[scale=0.45]{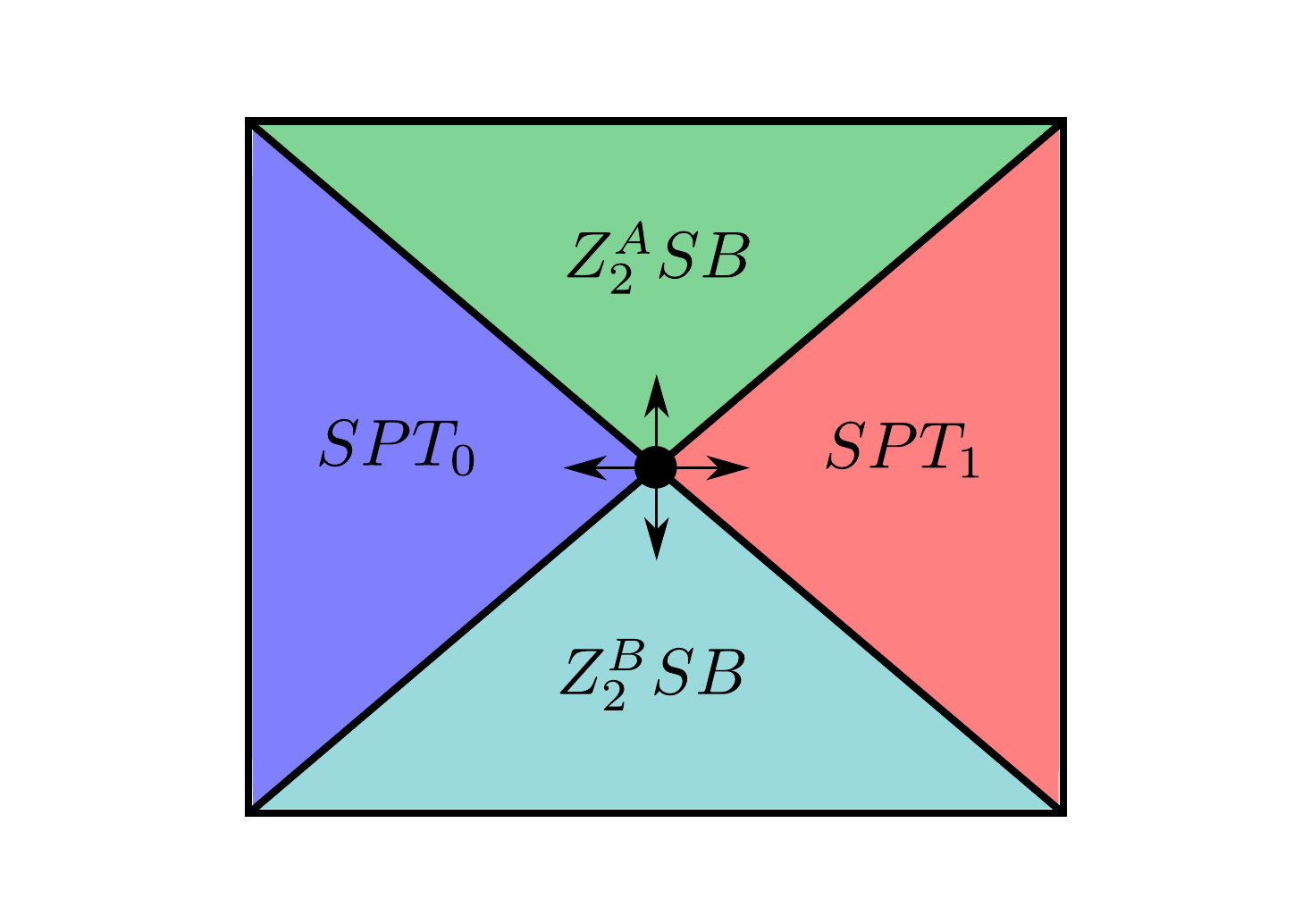}
\caption{(Color online) A schematic phase diagram near the $Z_2\times Z_2$ SPT critical point (the black point). The vertical and horizontal arrows correspond to perturbations associated with the two relevant operators found in section \ref{appdx:z2scaldim}. The relevant perturbation represented by the horizontal arrows drives the transition between the trivial SPT(blue) and the non-trivial SPT (red). The perturbation represented by the vertical arrows drives a Landau forbidden transition between spontaneous symmetry breaking (SB) phases where different $Z_2$ symmetries are broken in the two different phases (turquoise and green).}
\label{fig:znznphase}
\end{figure}

The operator that drives the SPT phase transition must be (1) relevant, (2) translational invariant and (3) invariant under $Z_2\times Z_2$. In \Eq{z2z2z} the only term that contains operators (there are two such operators due to the multiplicity 2) satisfy these conditions is $2|\chi_{I}\chi_{\e}|^2$. The scaling dimension of $(I\e)(\bar{I\e})$ is $h+\bar{h}=1<2$ hence it is relevant. The momentum of this operator is $h-\bar{h}=0$ hence is translation invariant. Moreover according to Table \ref{tab:z2charges} there operators are invariant under $Z_2\times Z_2$. 
It turns out that one of these two relevant operators drives a symmetry breaking transition while the other drives the SPT transition (See \Fig{fig:znznphase}). 
From the scaling dimension $h+\bar{h}=1$ we predict the gap exponent to be $\frac{1}{2-1}=1$.

\section{The modular invariant partition function and the primary scaling operators of the orbifold critical $Z_3\times Z_3$ clock model}\label{orbifoldz3xz3}

\subsection{Review of modular invariant partition function for the 3 states Potts model\\}\label{appdx:3statespotts}

The construction of the orbifold partition function for the $Z_3\times Z_3$ case closely mirrors the $Z_2\times Z_2$ case. But instead of two critical Ising chains, we now have two critical Potts chains. We first review the known results for the modular invariant $Z_3$ clock model (equivalent to the 3-state Potts model). The 3-state Potts model shows an order-disorder phase transition. At the critical point the Hamiltonian is given by 
\begin{align*}
H_{Potts}=-\sum_i \left[M_i + R^{\dg}_i R_{i+1} + h.c. \right]
\end{align*}
where $R_j=1,\eta_3,\eta_3^2$ ($\eta_3=e^{i2\pi/3}$) and $R_j M_k = \eta_3^{\delta_{jk}} M_k R_j$.
The conformal field theory for the critical 3-state Potts model belong to the well known ``minimal'' model  $\mathcal{M}(6,5)$\cite{Francesco2012,Dotsenko1984}. 
The central charge is 
\be
c={4\over 5}\label{cc}\ee
and the primary scaling operators are labeled by two pairs of indices $(r,s)$ and $(r^\prime,s^\prime)$ each label the 
``holomorphic'' and the ``anti-holomorphic'' part of the operator. The range of these indices are given by $1\le s\le r\le 4$ and $1\le s^\prime\le r^\prime\le 4$. 
The scaling dimensions of the holomorphic and anti-holomorphic parts of these operators are given by
\be
h_{r,s}={(6r-5s)^2 {-1}\over 120},~~\bar{h}_{r^\prime,s^\prime}={(6r^\prime-5s^\prime)^2 {-1}\over 120}. \ee  It is easy to check that $h_{r,s}=h_{5-r,6-s}$ and $\bar{h}_{r^\prime,s^\prime}=\bar{h}_{5-r^\prime,6-s^\prime}$ hence there are 10 distinct primary fields in the holomorphic and anti-holomorphic sector each. \\

Requiring modular invariance \eqref{eq:modinv} for $c=4/5$ yields two possible such $M$'s: one with $M_{(r,s);(r^\prime,s^\prime)}=\delta_{(r,s),(r^\prime,s^\prime)}$ describing the ``tetra-critical Ising model'', and the other corresponds to the 3-state Potts model described by the following partition function\cite{Francesco2012}:
\begin{align}
Z^{\rm 3-Potts}(\t)=|\ch_{I}(\t)|^{2}+|\ch_{\e}(\t)|^{2}+2|\ch_{\psi}(\t)|^{2}+2|\ch_{\s}(\t)|^{2}, \label{eq:Zpotts}
\end{align}
where 
\begin{align}
&\ch_{I}:=\ch_{1,1}+\ch_{4,1}, ~\ch_{\e}:=\ch_{2,1}+\ch_{3,1}, ~\ch_{\psi}:=\ch_{4,3}, ~\ch_{\s}:=\ch_{3,3}\label{ch}
\end{align}
Note that out of the 10 possible primary operators in each holomorphic/anti-holomorphic sector only six of them contribute to the partition function. In addition, the diagonal combination of the $(3,3)$ and $(4,3)$ operators from each sector {appear} twice. The explicit form of $\ch_{(r,s)}$ is given by equation (8.15) of Ref.[\onlinecite{Francesco2012}].
The conformal dimensions of primary fields and their eigenvalues under the action of the $Z_3$ generator are summarized in table \ref{tab:pottsh}\cite{Francesco2012}.

\begin{table*}[h]
\centering
\caption{Conformal dimensions of the primary fields of the 3-states Potts model, and their phases under the transformation of the $Z_3$ generator.}
\vspace{0.1in}
\begin{tabular}{|c|c|c|c|c|c|c|}
\hline
$(r,s)$		&(1,1)		&(2,1)		&(3,1)		&(4,1)		&$(3,3)_{1,2}$			&$(4,3)_{1,2}$	 \\
\hline
$h_{(r,s)}$	&0			&2/5		&7/5		&3			&1/15			&2/3\\
\hline
$Z_3$		&1			&1			&1			&1			&($\eta_3, \bar{\eta}_3)$		&($\eta_3, \bar{\eta}_3)$\\
\hline
\end{tabular}
\label{tab:pottsh}
\end{table*}

\subsection{Constructing the orbifold partition function for the $Z_3\times Z_3$ critical theory}\label{appdx:z3orbifold}

With the brief review of the modular invariant partition function of the critical 3-state Potts model we are ready to 
construct the partition function for the orbifolded $Z_3\times Z_3$ model defined by of \Eq{orbifoldZ}: 
\be
Z_{Z_3\times Z_3}(\t)
=\frac{1}{3} \sum_{q_s=0}^2\sum_{q_\tau=0}^{2} Z^{\rm 3-Potts}_{q_s,q_\tau}(\t)\times Z^{\rm 3-Potts}_{q_s,q_\tau}(\t)
\label{ofp}\ee
$Z^{\rm 3-Potts}_{(01)}$ and $Z^{\rm 3-Potts}_{(02)}$ are given in Ref.[\onlinecite{Francesco2012}]. Using $Z^{\rm 3-Potts}_{q_s,q_{\tau}}(\t)=Z^{\rm 3-Potts}_{q_s,q_{\tau}q_s}(\t+1)=Z^{\rm 3-Potts}_{q_{\tau}^{-1},q_s}(-1/\t)$, we have
\be
&&Z^{\rm 3-Potts}_{(00)}(\t)=|\ch_{I}(\t)|^{2}+|\ch_{\e}(\t|^{2}+2|\ch_{\psi}(\t)|^{2}+2|\ch_{\s}(\t)|^{2}\nn
&&Z^{\rm 3-Potts}_{(01)}(\t)=|\ch_{I}(\t)|^{2}+|\ch_{\e}(\t)|^{2}+(\eta_3+\bar{\eta}_3)|\ch_{\psi}(\t)|^{2}+(\eta_3+\bar{\eta}_3)|\ch_{\s}(\t)|^{2}\nn
&&Z^{\rm 3-Potts}_{(02)}(\t)=Z^{\rm 3-Potts}_{(01)}(\t)\nn
&&Z^{\rm 3-Potts}_{(10)}(\t)=Z^{\rm 3-Potts}_{(01)}(-1/\t)=\mathcal{S}Z^{\rm 3-Potts}_{(01)}(\t)\nn
&&Z^{\rm 3-Potts}_{(20)}(\t)=\mathcal{S}Z^{\rm 3-Potts}_{(02)}(\t)\nn
&&Z^{\rm 3-Potts}_{(12)}(\t)=Z^{\rm 3-Potts}_{(10)}(\t+1)=\mathcal{T}Z^{\rm 3-Potts}_{(10)}(\t)\nn
&&Z^{\rm 3-Potts}_{(11)}(\t)=\mathcal{T}Z^{\rm 3-Potts}_{(12)}(\t)\nn
&&Z^{\rm 3-Potts}_{(21)}(\t)=\mathcal{T}Z^{\rm 3-Potts}_{(20)}(\t)\nn
&&Z^{\rm 3-Potts}_{(22)}(\t)=\mathcal{T}Z^{\rm 3-Potts}_{(21)}(\t)
\ee
Using the S,T matrices of the 3-states Potts model we can compute all these terms. Substituting the results into \Eq{ofp} we obtain the orbifolded $Z_3\times Z_3$ partition function: 
\begin{align}
Z_{Z_3\times Z_3}&=(|\ch_{I}|^{2}+|\ch_{\e}|^{2})^{2} +4(|\ch_{\psi}|^{2}+|\ch_{\s}|^{2})^{2} +4|\ch_{I}\bar{\ch}_{\psi}+\ch_{\e}\bar{\ch}_{\s}|^{2},\label{finalZ}
\end{align}
where the $\t$ dependence is suppressed.  When expanded {in terms of $\ch_{r,s}\bar{\ch}_{r^\prime,s^\prime}$ }the first term yields 64 terms (henceforth referred as group I terms). The second term yields 16 terms (group II terms). The third term yields 64 terms (group III terms).
Due to the prefactor of 4 in the last two terms of \Eq{finalZ}  group II and III terms appear with multiplicity 4. Thus there are in total 144 terms, each corresponds to  a primary scaling operator. In Table \ref{tab:levels} we give the first few energy ($h+\bar{h}$)  and momentum ($h-\bar{h}$) eigenvalues 

\begin{table*}[h]
\centering
\caption{The quantum numbers of the first few primary operators of the orbifold $Z_3\times Z_3$ CFT.}
\vspace{0.1 in}
\begin{tabular}{|c|c|c|c|c|}
\hline
$h+\bar{h}$		&$h-\bar{h}$	 &Multiplicity				&Terms in $Z_{Z_3\times Z_3}$		\rule{0pt}{2.5ex}			\\
\hline
0								&0			&1				&$|\ch_{I}|^{4}$			 \\
\hline
4/15								&0			&4			&$4|\ch_{\s}|^{4}$			\\
\hline
4/5									&0		&2				&$2|\ch_{I}|^{2}|\ch_{\e}|^{2}$\\
\hline
14/15								&0			&4			&$4|\ch_{\e}|^{2}|\ch_{\s}|^{2}$	\\
\hline
17/15								&$\pm 1$		&8		&$4(\bar{\ch}_{I}\bar{\ch}_{\s}\ch_{\psi}\ch_{\e}+\text{c.c})$				\\
\hline
4/15+1							&$\pm1$			&16			&$4|\ch_{\s}|^{4}$(due to the first descendants)				\\
\hline
4/3								&0				&4			&$4|\ch_{I}|^{2}|\ch_{\psi}|^{2}$	\\
\hline
22/15						&0				&8				&$8|\ch_{\s}|^{2}|\ch_{\psi}|^{2}$		\\
\hline
8/5								&0				&1			&$|\ch_{\e}|^{4}$				\\
\hline
\end{tabular}
\label{tab:levels}
\end{table*}

\subsection{Transformation properties under the action of $Z_3 \times Z_3$} \label{appdx:z3z3transform}

Following the same procedure in \ref{appdx:z2z2transform} we construct operators that project the Hilbert space into subspaces carrying various irreducible representation of $Z_3\times Z_3$ which is generated by $G_A=\widetilde{B}$ and $G_B=\prod_{i \in even} \widetilde{M}_i$. The projector into subspace with eigenvalues $(\eta_3^{a},\eta_3^{b})$ (here $\eta_3=e^{i 2\pi/3}$ and $a,b=0,1,2$) under the action of $G_A$ and $G_B$ is given by
\be
P_{ab}=\left(\frac{1+\eta_3^{-a}G_A+\eta_3^{a}G_A^{2}}{3}\right) \left(\frac{1+\eta_3^{-b}G_B+\eta_3^{b}G_B^{2}}{3}\right) \ee
Analogous to \Eq{z2partf} we filter out the Verma modules that transform according to this particular irreducible representation by computing
\begin{align}
P_{ab}Z_{Z_3\times Z_3}&:=\frac{1}{27}\sum_{q_\tau,q_s,\mu,\nu=0}^{2}\eta_3^{-a\mu-b\nu}\Big[\eta_3^{q_s\mu}(Z^{\rm 3-Potts}_{q_s,q_{\t}})(Z^{\rm 3-Potts}_{q_s,q_{\t}+\nu})\Big]
\label{partf}
\end{align}
\\

For example, \be P_{00}Z_{Z_3\times Z_3}=(|{\ch_{I}}|^{2}+|{\ch_{\e}}|^{2})^{2},\label{z3sya}\ee which means only group I transform as the identity representation of $Z_3 \times Z_3$. For other $P_{ab}$ the results are summarized in table \ref{tab:charges}

\begin{table*}[h]
\centering
\caption{Transformation properties of the contributing Verma modules in \Eq{finalZ} under the action of $G_A$ and $G_B$. For group II and group III, the quadruplet records the transformation properties of the multiplicity four Verma modules in \Eq{finalZ} .}
\vspace{0.1 in}
\begin{tabular}{|c|c|c|}
\hline
Group	&$G_A$	&$G_B$ \\
\hline
I		&1			&1\\
\hline
II		&$(\eta_3,\bar{\eta}_3,1,1)$		&$(1,1,\eta_3,\bar{\eta}_3)$\\
\hline
III		&$(\eta_3,\eta_3,\bar{\eta}_3,\bar{\eta}_3)$		&$(\eta_3,\bar{\eta}_3,\eta_3,\bar{\eta}_3)$\\
\hline
\end{tabular}
\label{tab:charges}
\end{table*}

\subsection{Scaling Dimension for the operator driving the $Z_3\times Z_3$ SPT transition}\label{appdx:scaldim}

From Table \ref{tab:levels} and \Eq{z3sya} it is seen that the translation-invariant (\ie $h-\bar{h}=0$), relevant(\ie $h+\bar{h}<2$), $Z_3\times Z_3$ invariant 
operators either have scaling dimensions $4/5$ or $8/5$. Through a comparison with the numerical result for the gap exponent in section 9 of the main text, we identify one of the operators with scaling dimension $4/5$ as responsible for the {opening of} the energy gap in the SPT phase transition. The predicted gap exponent is $\frac{1}{2-4/5}=5/6$ which agrees reasonably well with the numerical gap exponent. Moreover similar to the $Z_2\times Z_2$ case there are two operators with the same scaling dimension (4/5). Again one of these operators drives a symmetry breaking transition while the other drives the SPT transition, hence the phase diagram is similar to \Fig{fig:znznphase}.

\section{The modular invariant partition function and the primary scaling operators of the orbifold critical $Z_4\times Z_4$ clock model}\label{orbifoldz4xz4}

\subsection{Review of modular invariant partition function for the $Z_4$ clock model}\label{appdx:z4 clock}

The $Z_4$  clock model undergoes an order-disorder transition. The Hamiltonian at criticality between the ordered is given by
\begin{align}
H_{Z_4}=-\sum_{i=1}^{N} \left[M_i + R^{\dg}_i R_{i+1} + h.c. \right] \label{eq:z4H}
\end{align}
where $R_j=1,\eta_4,\eta_4^2,\eta_4^3$ where $\eta_4=e^{i 2\pi/4}$, and $R_j M_k = \eta_4^{\delta_{jk}} M_k R_j$. With periodic boundary condition, $R_{N+1}=R_{1}$, it can be exactly mapped onto two decoupled periodic Ising chains\cite{Li2015} as follows. For the $Z_4$ clock model the Hilbert space for each site $j$ is 4-dimensional. In the following, we shall regard this 4-dimensional Hilbert space as the tensor product of two 2-dimensional Hilbert spaces associated with site $2j-1$ and $2j$. We then view each of the 2-dimensional space as the Hilbert space of an Ising spin. In this way the $Z_4$ clock model with $N$ sites can be viewed as an Ising model with $2N$ sites.\\

More explicitly, under the unitary transformation $U=\prod_i U_i$, where
\begin{align*}
U_i&=\bpm 
0 & 1 & 0 & 0\\
1 & 0 & 0 & 0\\
0 & 0 & 1 & 0\\
0 & 0 & 0 & 1 \\
\epm, 
\end{align*}
\Eq{eq:z4H} becomes
\begin{align}
U^{\dg}H_{Z_4}U & =-\sum_{i=1}^{N}\left(X_{2i-1}+Z_{2i-1}Z_{2i+1}\right)-\sum_{i=1}^{N}\left(X_{2i}+Z_{2i}Z_{2i+2}\right) \label{eq:z4toz2}\\
&=H^{\rm Ising}_{odd}+H^{\rm Ising}_{even} \nonumber
\end{align}
where $X_i$ and $Z_i$ denote the $2\times 2$ Pauli matrices $\s_i^x$ and $\s_i^z$. Thus the partition function of the $Z_4$ clock model under periodic boundary condition is given by
\begin{align*}
Z^{\rm 4-clock}_{(0,0)}(\tau)=Z^{\rm Ising}_{(0,0)}(\tau)\times Z^{\rm Ising}_{(0,0)}(\tau)
\end{align*}
The fact that Ising model has central charge $c=1/2$ implies the central charge of the critical $Z_4$ clock model to be  $1/2+1/2=1$.\\

CFT with $c=1$ has infinitely many Verma modules\cite{Feigin1983}. The scaling dimension of the primary fields, which can take any non-negative values, is parametrized by $h=x^{2}/4$ where $x$ is a non-negative real number. The characters associated with these Verma modules are given\cite{Runkel2001} by
\begin{align}
\chi_h(q)=
\begin{cases}
\frac{1}{\eta(q)}q^{x^{2}/4},&~~\text{for}~x\notin\mathbb{Z}\\
\frac{1}{\eta(q)}\left(q^{x^{2}/4}-q^{(x+2)^{2}/4}\right),&~~\text{for}~x\in\mathbb{Z} 
\end{cases}
\label{eq:z4irrchar}
\end{align}
where \be\eta(q)=q^{1/24}\prod_{n=1}^{\infty}\left(1-q^{n}\right).\ee \\

Because later on we shall perform orbifolding it is necessary to consider the $Z_4$ clock model under twisted spatial boundary condition.
With the spatial boundary condition twisted by the $Z_4$ generator, i.e., $R_{N+1}=\eta_4 R_{1}$, 
the last two terms, namely $Z_{2N}Z_{2}+Z_{2N-1}Z_{1}$ in \Eq{eq:z4toz2}, are replaced by
\begin{align*}
Z_{2N}Z_{2}+Z_{2N-1}Z_{1}\ra Z_{2N}Z_{1}-Z_{2N-1}Z_{2}
\end{align*}
\begin{figure}[h!]
\centering
\includegraphics[scale=0.55]{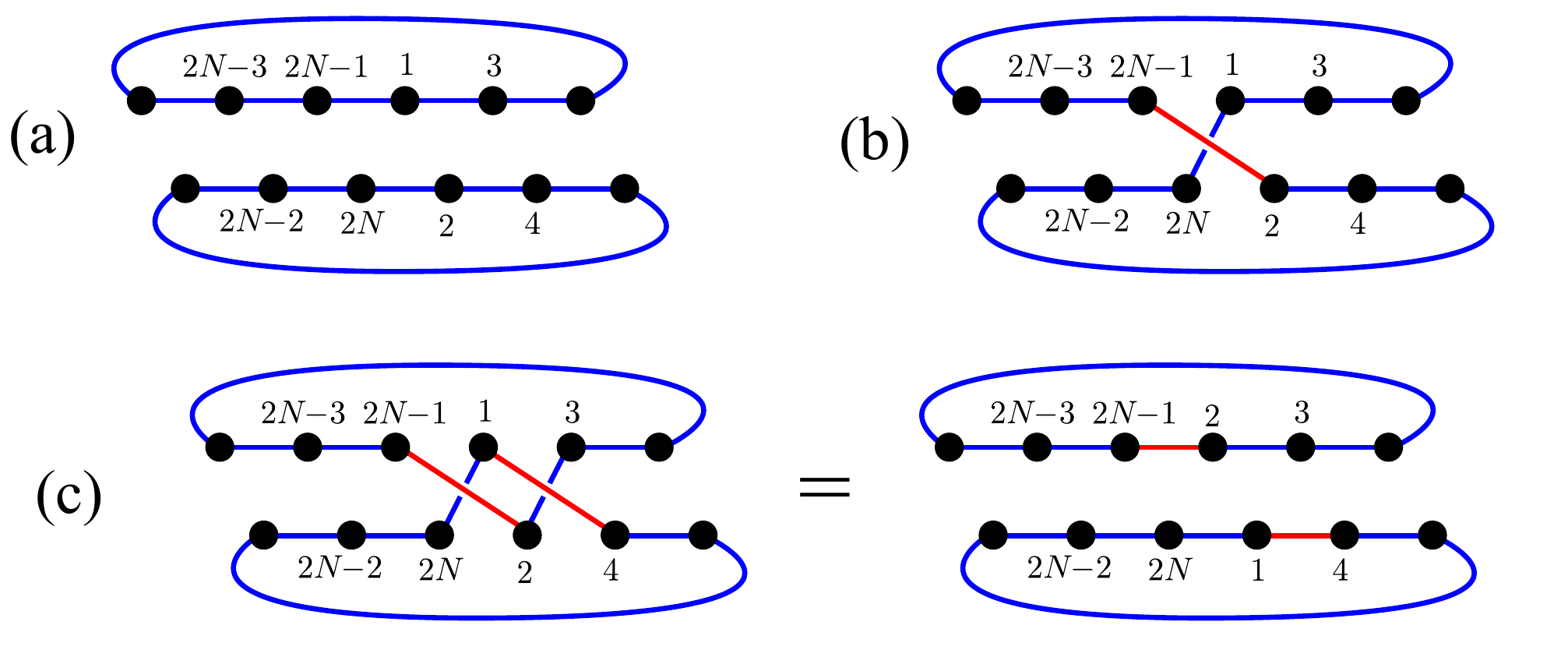}
\caption{(Color online) The mapping of $Z_4$-clock model with different spatially twisted boundary conditions to Ising models. Each black dot represents the $X_i$ term in the Hamiltonian and each blue bond represents the term $Z_iZ_j$ (an antiferromagnetic bond). The red bond represents $-Z_iZ_j$. (a)With periodic boundary condition, the $Z_4$ clock model maps to two decoupled Ising chains. (b)When the  boundary condition is twisted by a $Z_4$ generator, the $Z_4$ clock model maps to a single Ising chain twice as long with one antiferromagnetic bond. (c)When the boundary condition is twisted by the square of the $Z_4$ generator, the $Z_4$ clock model maps to two decoupled Ising chains, each having an antiferromagnetic bond.}
\label{fig:z4toz2}
\end{figure}
In the language of Ising model, the above replacement creates an overpass connecting the even chain to the odd chain and  
a sign change of one bond (the red bond in \Fig{fig:z4toz2}(b)). Thus we arrive at an Ising chain twice as long and with the spatial boundary condition twisted by the $Z_2$ generator. As a result 
\be
Z^{\rm 4-clock}_{(1,0)}(\tau)=Z^{\rm Ising}_{(1,0)}(\tau/2).\label{tw4}
\ee
The reason the modular parameter of the Ising partition function is half that of the $Z_4$ clock partition function is because the Ising chain has twice the length in the spatial direction.  
The same argument applies if the boundary is twisted by the inverse of the $Z_4$ generator ($R_{N+1}=\eta_4^3 R_{1}$) instead, i.e., 
\be
Z^{\rm 4-clock}_{(3,0)}(\tau)=Z^{\rm Ising}_{(1,0)}(\tau/2).
\ee

Similarly, when the spatial direction is $R_{N+1}=\eta_4^2 R_{1}$, the Hamiltonian of the Ising model becomes that of two decoupled Ising chain each having a sign-flipped bond equivalent to the $Z_2$ twisted boundary condition (See \Fig{fig:z4toz2}(c)).  The resulting partition function is given by
\begin{align*}
Z^{\rm 4-clock}_{(2,0)}(\tau)=Z^{\rm Ising}_{(1,0)}(\tau)\times Z^{\rm Ising}_{(1,0)}(\tau)
\end{align*}

Using the known $S$ and $T$ matrices for the Ising model, other $Z^{\rm 4-clock}_{(q_s,q_t)}(\tau)$ can be determined
\begin{align}
Z^{\rm 4-clock}_{(0,1)}(\tau)&=Z^{\rm 4-clock}_{(0,3)}(\tau)=\mathcal{S}Z^{\rm 4-clock}_{(1,0)}(\tau)=Z^{\rm Ising}_{(0,1)}(2\tau)\nn
Z^{\rm 4-clock}_{(1,3)}(\tau)&=Z^{\rm 4-clock}_{(3,1)}(\tau)=\mathcal{T}Z^{\rm 4-clock}_{(3,0)}(\tau)=Z^{\rm Ising}_{(1,0)}(\tau/2+1/2)\nn
Z^{\rm 4-clock}_{(1,1)}(\tau)&=Z^{\rm 4-clock}_{(3,3)}(\tau)=\mathcal{S}Z^{\rm 4-clock}_{(1,3)}(\tau)=Z^{\rm Ising}_{(1,1)}(\tau/2+1/2)\nn
Z^{\rm 4-clock}_{(1,2)}(\tau)&=Z^{\rm 4-clock}_{(3,2)}(\tau)=\mathcal{T}Z^{\rm 4-clock}_{(1,3)}(\tau)=Z^{\rm Ising}_{(1,1)}(\tau/2)\nn
Z^{\rm 4-clock}_{(2,1)}(\tau)&=Z^{\rm 4-clock}_{(2,3)}(\tau)=\mathcal{S}Z^{\rm 4-clock}_{(1,2)}(\tau)=Z^{\rm Ising}_{(1,1)}(2\tau)\nn
Z^{\rm 4-clock}_{(0,2)}(\tau)&=\mathcal{S}Z^{\rm 4-clock}_{(2,0)}(\tau)=Z^{\rm Ising}_{(0,1)}(\tau)\times Z^{\rm Ising}_{(0,1)}(\tau)\nn
Z^{\rm 4-clock}_{(2,2)}(\tau)&=\mathcal{T}Z^{\rm 4-clock}_{(2,0)}(\tau)=Z^{\rm Ising}_{(1,1)}(\tau)\times Z^{\rm Ising}_{(1,1)}(\tau)
\label{eq:z4partitions}
\end{align}

\subsection{Orbifold partition function for the critical $Z_4\times Z_4$ CFT}
Using these result and \Eq{orbifoldZ} we can calculate the orbifolded $Z_4\times Z_4$ partition function
\begin{align}
Z_{Z_4\times Z_4}(\t)
&=\frac{1}{4} \sum_{q_s=0}^3\sum_{q_\tau=0}^{3} Z^{\rm 4-clock}_{q_s,q_\tau}(\t)\times Z^{\rm 4-clock}_{q_s,q_\tau}(\t)\nn
&=\left(|\chi_{I}|^2+|\chi_{J}|^2+2|\chi_{Y}|^2+|\chi_{\e}|^2+\bar{\chi}_{\alpha}\chi_{\beta}+\bar{\chi}_{\beta}\chi_{\alpha}\right)^2\nn
&+\left(\bar{\chi}_{I}\chi_{J}+\bar{\chi}_{J}\chi_{I}+2|\chi_{Y}|^2+|\chi_{\e}|^2+\bar{\chi}_{\alpha}\chi_{\beta}+\bar{\chi}_{\beta}\chi_{\alpha}\right)^2\nn &+2\left(|\chi_{\alpha}|^2+|\chi_{\beta}|^2+|\chi_{\e}|^2+\bar{\chi}_{Y}\left(\chi_{I}+\chi_{J}\right)+\left(\bar{\chi}_{I}+\bar{\chi}_{J}\right)\chi_{Y}\right)^2\nn
&+4\left(|\chi_{\s}|^2+|\chi_{\t}|^2\right)^2+4\left(\bar{\chi}_{\s}\chi_{\t}+\bar{\chi}_{\t}\chi_{\s}\right)^2\nn
&+4|\bar{\chi}_{O_1}\chi_{O_3}+\bar{\chi}_{O_3}\chi_{O_2}+\bar{\chi}_{O_2}\chi_{O_4}+\bar{\chi}_{O_4}\chi_{O_1}|^2,
\label{eq:z4z4Z}
\end{align}
where
\begin{align}
\chi_{I}&=
\ch_0+\sum_{n>0}\left(\ch_{8n^2}+\ch_{4n^2}\right)\nn
\chi_{J}&=
\sum_{n>0}\left(\ch_{8n^2}+\ch_{(2n-1)^2}\right)\nn
\chi_{Y}&=
\sum_{n>0}\ch_{2(2n-1)^2};~
\chi_{\e}=
\sum_{n}\ch_{(4n+1)^2/2}\nn
\chi_{\alpha}&=
\sum_{n}\ch_{(8n+1)^2/8};~
\chi_{\beta}=
\sum_{n}\ch_{(8n+3)^2/8}\nn
\chi_{\sigma}&=
\sum_{n}\ch_{(8n+1)^2/16};~
\chi_{\tau}=
\sum_{n}\ch_{(8n+3)^2/16}\nn
\chi_{O_1}&=
\sum_{n}\ch_{(16n+1)^2/32};~
\chi_{O_2}=
\sum_{n}\ch_{(16n+7)^2/32}\nn
\chi_{O_3}&=
\sum_{n}\ch_{(16n+3)^2/32};~
\chi_{O_4}=
\sum_{n}\ch_{(16n+5)^2/32}.
\label{eq:z4extendedchar}
\end{align}
The $\ch_{h}$ in the above equations are given by \Eq{eq:z4irrchar}.
The scaling dimensions of the highest weight states associated with the Verma modules that generate these $\chi_h$ are summarized in Table \ref{tab:4sh}.
\begin{table*}[h]
\centering
\caption{Scaling dimensions of the extended primary fields of the $Z_4$ clock model.}
\vspace{0.1in}
\begin{tabular}{|c|c|c|c|c|c|c|c|c|c|c|c|c|}
\hline
field		&$I$		&$J$		&$Y$		&$\e$		&$\alpha$			&$\beta$	&$\s$	&$\t$	&$O_1$	&$O_2$	&$O_3$	&$O_4$	 \\
\hline
$h$	&$0$			&$1$		&$2$		&${1/2}$	&${1/8}$	&${9/8}$	&${1/16}$	&${9/16}$	&${1/32}$	&${49/32}$		&${9/32}$			&${25/32}$\\
\hline
\end{tabular}
\label{tab:4sh}
\end{table*}

Let's refer to the six terms in \Eq{eq:z4z4Z} as Group I, II, III, IV, V, and VI respectively. Due to the prefactor of 2, Group III elements appear in doublets. Due to the prefactor of 4, Group IV, V and VI elements appear with multiplicity 2. In Table \ref{tab:z4levels} we list the first few primary fields with scaling dimension $h+\bar{h}<2$ and their quantum numbers.  
\begin{table*}[h]
\centering
\caption{The quantum numbers of the first few low scaling dimension primary operators of the orbifold $Z_4\times Z_4$ CFT.}
\vspace{0.1 in}
\begin{tabular}{|c|c|c|c|c|}
\hline
$h+\bar{h}$		&$h-\bar{h}$	 &Multiplicity				&Terms in $Z_{Z_4\times Z_4}$		\rule{0pt}{2.5ex}			\\
\hline
0				&0			&1			&$|\ch_{I}|^{4}$			 \\
\hline
1/4				&0			&4			&$4|\ch_{\s}|^{4}$			\\
\hline
1/2				&0			&2			&$2|\ch_{\alpha}|^{4}$	\\
\hline
5/8				&0			&4			&$4|\ch_{O_1}\ch_{O_3}|^{2}$	\\
\hline
1				&0			&2			&$2|\ch_{I}\ch_{\e}|^2$				\\
\hline
9/8				&$\pm1$		&8			&$4\left(\bar{\ch}_{O_1}^2\ch_{O_3}\ch_{O_4}+\text{c.c.}\right)$		\\
\hline
5/4				&0			&16+4		&$16|\ch_{\s}\ch_{\t}|^{2}+4|\ch_{\alpha}\ch_{\e}|^{2}$		\\
\hline
5/4				&$\pm1$		&8+4		&$4\left(\bar{\ch}_{\s}\ch_{\t}\right)^2+2|\ch_{I}|^{2}\bar{\ch}_{\alpha}\ch_{\beta}+\text{c.c.}$		\\
\hline
5/4				&$\pm1$		&16		&$4|\ch_{\s}|^{4}$(first descendants)		\\
\hline
3/2				&$\pm1$		&8		&$2|\ch_{\alpha}|^{4}$(first descendants)		\\
\hline
13/8			&0			&4			&$4|\ch_{O_1}\ch_{O_4}|^{2}$		\\
\hline
13/8			&$\pm1$		&16		&$4|\ch_{O_1}\ch_{O_3}|^{2}$(first descendants)	\\
\hline
\end{tabular}
\label{tab:z4levels}
\end{table*}\\

\subsection{Transformation properties under the action of $Z_4 \times Z_4$} 
Similar to section \ref{appdx:z2z2transform} and \ref{appdx:z3z3transform} we resolve the Verma modules that generate the partition function in \Eq{eq:z4z4Z} into different irreducible representation spaces of $Z_4\times Z_4$. As done in previous sections we 
construct the symmetry projection operators
\begin{align}
P_{ab}Z_{Z_4\times Z_4}&:=
&\frac{1}{64}\sum_{q_\tau,q_s=0}^{3}\sum_{\mu,\nu=0}^{3}\eta_4^{-a\mu-b\nu+q_s\mu}\Big[Z^{\rm 4-clock}_{q_s,q_{\t}}Z^{\rm 4-clock}_{q_s,q_{\t}+\nu}\Big]
\end{align}
The results are summarized in Table \ref{tab:z4z4charges}.
\begin{table*}[h]
\centering
\caption{Transformation properties of the contributing Verma modules in \Eq{eq:z4z4Z} under the action of $G_A$ and $G_B$. For Group III to VI, the multiplet records the transformation properties of the corresponding degenerate Verma modules in \Eq{eq:z4z4Z}.}
\vspace{0.1 in}
\begin{tabular}{|c|c|c|}
\hline
Group	&$G_A$	&$G_B$ \\
\hline
I		&1			&1\\
\hline
II		&$-1$			&$-1$\\
\hline
III		&$(1,-1)$		&$(-1,1)$\\
\hline
IV		&$(1,1,\eta_4,\bar{\eta}_4)$		&$(\eta_4,\bar{\eta}_4,1,1)$\\
\hline
V		&$(-1,-1,\eta_4,\bar{\eta}_4)$		&$(\eta_4,\bar{\eta}_4,-1,-1)$\\
\hline
VI		&$(\eta_4,\eta_4,\bar{\eta}_4,\bar{\eta}_4)$		&$(\eta_4,\bar{\eta}_4,\eta_4,\bar{\eta}_4)$\\
\hline
\end{tabular}
\label{tab:z4z4charges}
\end{table*}
\\

The term $2|\ch_{I}\ch_{\e}|^2$ in \Eq{eq:z4z4Z} yields two primary fields with scaling dimension $h+\bar{h}=1$ (hence are relevant) and are invariant under $Z_4\times Z_4$ and translation. Hence they are qualified as the gap generating operator. The gap exponent is $\frac{1}{2-1}=1$. Similar to the $Z_2\times Z_2$ and $Z_3\times Z_3$ cases there are two operators with the same scaling dimension (1). As the above two cases one of these operators drives a symmetry breaking transition while the other drives the SPT transition, hence the phase diagram is similar to \Fig{fig:znznphase}. 

\section{Some details of the density matrix renormalization group calculations}\label{DMRG}

\subsection{The truncation error estimate}

We determine the ground state phase diagram and properties of the model Hamiltonian in \Eq{eq:hlambda} by extensive and highly accurate density-matrix renormalization group \cite{White1992DMRG} (DMRG) calculations. We consider both periodic (PBC) and  open (OBC) boundary conditions. Careful study of the dependence on the finite system sizes enables the extrapolation to the thermodynamic limit. For OBC, we keep up to $m = 1000$ states in the DMRG block with around 24 sweeps to get converged results. The truncation error is estimated to be no bigger than $\epsilon=5\times 10^{-9}$. 
For PBC, we keep up to $m=1100$ states with around 60 sweeps for converged results. In this case, the truncation error is of the order $\epsilon=10^{-5}$.\\

\subsection{The entanglement entropy}

For conformal invariant system in one dimension, the central charge can be extracted by fitting the von Neumann entanglement entropy to the following analytical form \cite{CFTCC2004}
 \begin{eqnarray}
 S(x) = \frac{c}{3\eta} \ln(x) + {\rm constant}. \label{Eq:EEFit}
 \end{eqnarray}
Here $\eta=1 (2)$ for the periodic (open) boundary condition, respectively. The parameter $x$ is given by $x =\frac{\eta N}{l} \sin(\frac{\pi l}{N})$ 
for a cut dividing the chain into segments of length $l$ and $N-l$. 
For each system size under both OBC and PBC, we first calculate the entanglement entropy 
by keeping a fixed number of states $m$, hence yielding finite truncation error $\epsilon$. We then perform systematic $m$ dependence study which allows us to extrapolate to the $\epsilon=0$ limit. For each system size the resulting entanglement entropy is fit to \Eq{Eq:EEFit} to generate the data shown in \Fig{fig:z3c}. This result enables us to estimate  the central charge to be $c=\frac{8}{5}$. The exponent for the energy gap is obtained in a similar way.\\


\section{The on-site global symmetry of conformal field theories}\label{appdx:symm}

In order to determine which CFT can describe the critical points between bosonic SPT phases, it is important to understand the on-site symmetries of CFTs. This is because the critical theory should at least contain the protection symmetry (which is on-site for bosonic SPTS) of the SPT phases on either side. In Ref.\onlinecite{Pasquier1987} it is shown that a particular type of lattice models (the ``RSOS models'') reproduce the minimal model CFTs in the continuum limit. Moreover, the symmetry of such lattice model is related to that of the Dynkin diagrams which are used to classify the modular invariant partition functions\cite{Pasquier1987a,Pasquier1987}. However this elegant result does not answer the question whether the continuum theory has emergent symmetry beyond that of the lattice model. In this appendix we briefly review the results of Ref.\onlinecite{ZUBER1998,Ruelle1998} which answers this question.  
\\

\begin{figure}[h!]
\centering
\includegraphics[scale=0.4]{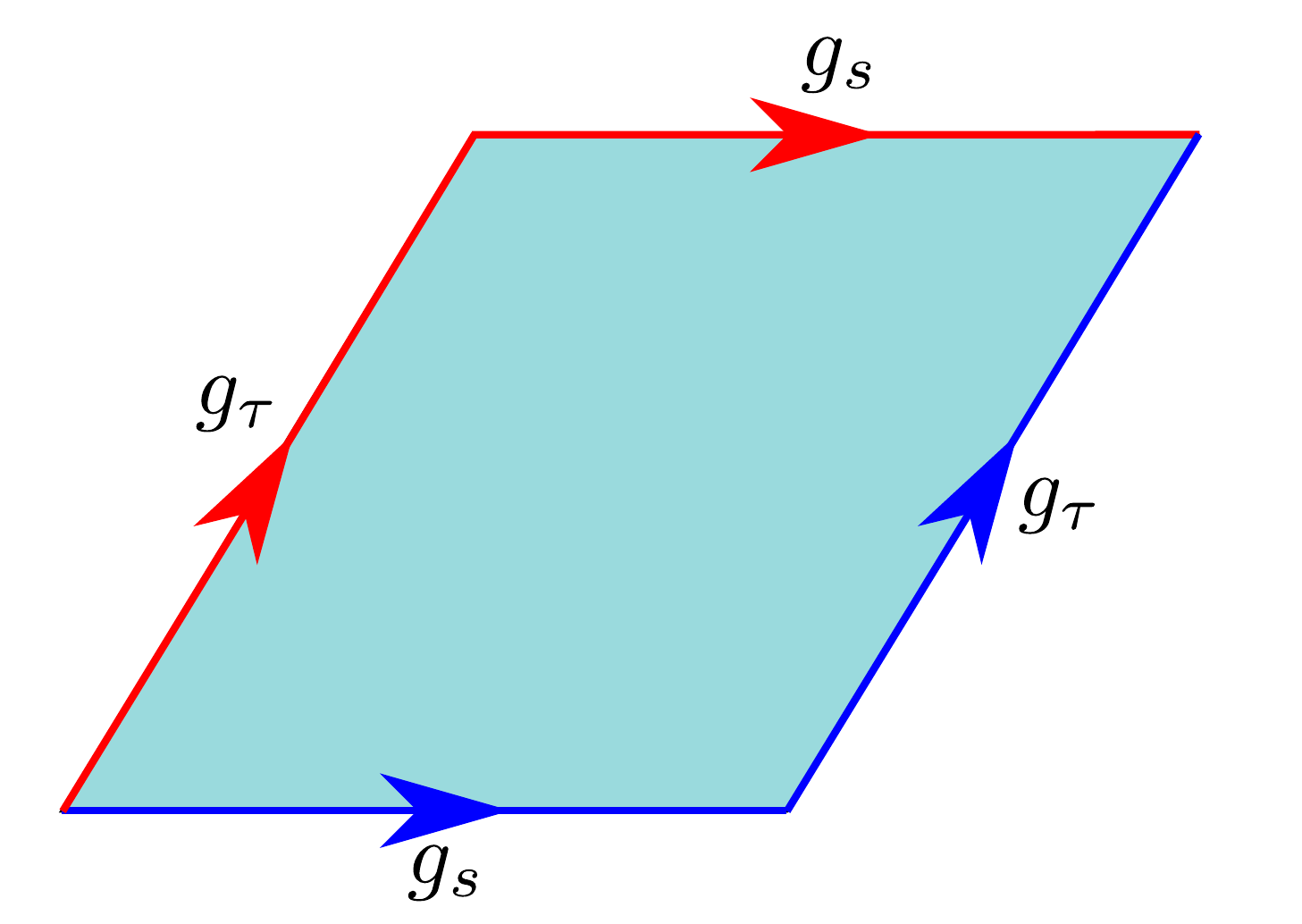}
\caption{(Color online) The space-time torus with spatial and temporal boundary condition twisted by group elements $g_s$ and $g_{\t}$. 
The path in red picks up the group element $g_{\t}g_s$, while the path in blue picks up the group element $g_{s}g_{\t}$. Since the path in red can be deformed into the path in blue, $g_s$ and $g_\tau$ need to commute so that the boundary condition is self-consistent.}
\label{twistcommute}
\end{figure}

The key idea of Ref.\onlinecite{Ruelle1998} is the following. Let's assume the CFT in question has an on-site symmetry group $G$. This means $G$ commutes with the Virasoro algebra hence each Verma module must carry an irreducible representation of $G$. Let $Z_m$ be any abelian subgroup of $G$. We can use $Z_m$ to perform orbifolding. (Note that in order for the space and time symmetry twists to be consistent with each other on a torus the respective elements we use to twist the space and time boundary conditions must commute 
(see figure \ref{twistcommute}). Moreover, if the $Z_m$ irreducible representations are correctly assigned to the Verma modules the resulting orbifolded partition is modular invariant. Therefore to detect whether an on-site symmetry group contains $Z_m$ as an abelian subgroup we just need to see whether it possible to assign $Z_m$ irreducible representations to the Verma modules so that after orbifolding the partition function is modular invariant. For discrete groups after knowing all abelian subgroups we can reconstruct the total group $G$. This is essentially the strategy followed by Ref.\onlinecite{Ruelle1998}. \\ 

More explicitly, let the Hilbert space consistent with a spatial boundary condition involving a twist generated by $\rho^{g_s}$  ($\rho$ is the generator of certain abelian subgroup $Z_m$ and $g_s=0,...,m-1$)

\be
\mathcal{H}^{(g_s)}=\oplus_{i,j} \oplus_{k=1}^{\mathcal{M}_{ij}^{(g_s)}} (\mathcal{V}_i \otimes \bar{\mathcal{V}}_j)_k
\label{hsp}
\ee
\noindent where $\mathcal{V}_i$ ($\bar{\mathcal{V}}_i$) is the $i$th Verma module in the holomorphic (anti-holomorphic) sector and $\mathcal{M}_{ij}^{(g_s)}$ is a non-negative integer labeling the multiplicity of the $\mathcal{V}_i \otimes \bar{\mathcal{V}}_j$ modules. Moreover, for the CFT to have a unique ground state, we require the vacuum module ($i=1$) only shows up once in the periodic sector, i.e.,  $\mathcal{M}_{11}^{(g_s)}=\delta_{0,g_s}$.\\

Next we assign irreducible representation to the Verma modules:
\be
\rho^{g_{\tau}} (\mathcal{V}_i \otimes \bar{\mathcal{V}}_j)_k = \eta_m^{Q(g_{\t};g_s,i,j,k)} (\mathcal{V}_i \otimes \bar{\mathcal{V}}_j)_k
\label{symc}
\ee
where $g_\tau=0,...,m-1$, $\eta_m=e^{\frac{i 2 \pi}{m}}$ and $Q(g_{\t};g_s,i,j,k)\in {0,...,m-1}$ is called ``symmetry charge'' in Ref.\onlinecite{Ruelle1998}. Combine \Eq{hsp} and \Eq{symc} we obtain the following space-time boundary twisted partition function on a torus with modular parameter $\tau$
\be
Z_{g_s,g_{\tau}}(\tau)=&Tr_{\mathcal{H}^{(g_s)}}(q^{L_0-c/24} \bar{q}^{\bar{L}_0-c/24} q_{\tau}) 
		=& \sum_{i,j} \left[ \sum_{k=1}^{\mathcal{M}_{ij}^{(g_s)}}   \zeta_N^{Q(g_{\tau};g_s,i,j,k)} \chi_i(\tau) \bar{\chi}_j(\tau) \right]\nn\label{zst}
\ee

\subsection{The consistency conditions}

So far the abelian subgroup $Z_m$ as well as $\mathcal{M}_{ij}^{(g_s)}$ and $Q(g_{\t};g_s,i,j,k)$ are unknown. They need to be determined subjected to the following consistency conditions. (1) When there is no spatial boundary condition twist the Hilbert space in \Eq{hsp} must return to that of the periodic boundary condition. Moreover in the case
where there is also no time boundary condition twist the partition function must agree with the modular invariant 
partition function $Z_{0,0}(\tau)$. (2) The $Z_{g_s,g_{\t}}(\tau)$ in \Eq{zst} must transform under the generators (S and T) of the modular transformation as  (see \Fig{fig:STtau}):
\begin{align*}
Z_{g_s,g_{\tau}}(\t)=Z_{g_s,g_{\tau}g_s}(\t+1)=Z_{g_{\tau}^{-1},g_s}(-1/\t)
\end{align*}
(3) $\mathcal{M}_{11}^{(g_s)}=\delta_{0,g_s}$, $\mathcal{M}_{ij}^{(g_s)}=$ non-negative integer, and $Q(g_{\t};g_s,i,j,k)=0,...,m-1$. (1)-(3) pose strong constraints on the possible abelian subgroup $Z_m$ and the allowed assignment of the irreducible representations (i.e. $Q(g_{\t};g_s,i,j,k)$) to each Verma module. \\

\subsection{The on-site symmetry of minimal models}
 
Under constants (1)-(3) in the previous subsection  Ref.\onlinecite{Ruelle1998} solved the possible abelian subgroups and their symmetry representations for the all minimal models. By patching these abelian subgroups together the author reached the following conclusion: the on-site symmetries of the unitary minimal models are exactly the same as those predicted by the lattice RSOS models \cite{Pasquier1987}. Hence there is no emergent symmetry! Thus, for most of the unitary minimal models the  symmetry is $Z_2$. The only exceptions are 3-states Potts and tri-critical 3-state Potts models where the symmetry is $S_3$. Finally for the minimal model labeled by  $E_7,E_8$, where there is no symmetry.
\\

{\noindent\bf Acknowledgement} \\

\noindent We thank John Cardy, Geoffrey Lee, Yuan-Ming Lu, Shinsei Ryu and Jian Zhou for very helpful discussions.
This work was primarily funded by the U.S. Department of Energy (DOE), Office of Science, Office of Basic Energy Sciences (BES), Materials Sciences and Engineering Division under Contract no DE-AC02-05CH11231 within the Theory of Materials Program (KC 2301).
Work by H.C. Jiang was supported by the U.S. Department of Energy (DOE), Office of Science, Office of Basic Energy Sciences, Division of Materials Sciences and Engineering, under Contract No. DE-AC02-76SF00515. Analytic works were performed at Berkeley. Parts of the computing for this project was performed on the Sherlock cluster. We would like to thank Stanford University and the Stanford Research Computing Center for providing computational resources and support that have contributed to these research results.

\B{\nocite{Cho2016}}

\bibliographystyle{ieeetr}
\bibliography{bibs}

\end{document}